\documentclass[twocolumn]{aastex63}

\usepackage[utf8]{inputenc}
\usepackage{amsmath}
\usepackage{tabularx}
\usepackage{amssymb}
\usepackage{lineno}
\usepackage{nameref}
\usepackage{array,multirow,makecell}
\usepackage{graphicx}
\usepackage{epstopdf}

\submitjournal{ApJ}

\graphicspath{{./}{figures/}}

\begin{document}

\title{MULTI WAVELENGTH STUDY OF 4C+28.07}

\author[0000-0002-9526-0870]{Avik Kumar Das}
\affiliation{Raman Research Institute, C. V. Raman Avenue, Sadashivnagar, Bangalore: 560080, India}

\author[0000-0002-1173-7310]{Raj Prince}
\affiliation{Center for Theoretical Physics, Polish Academy of Sciences, Al.Lotnikow 32/46, 02-668, Warsaw, Poland}

\author[0000-0002-1188-7503]{Nayantara Gupta}
\affiliation{Raman Research Institute, C. V. Raman Avenue, Sadashivnagar, Bangalore: 560080, India}

\email{avikdas@rri.res.in}


\begin{abstract}

4C+28.07 is a $\gamma$-ray Flat Spectrum Radio Quasar (FSRQ) type source. It is often monitored at different frequencies, though long term multi-wavelength data of this source have not been modelled in detail before. We have analyzed $\sim 12$ years (Aug, 2008 - May, 2020) of Fermi-LAT data with a binning of 10 days time scale and observed three distinctive flaring states. Each flaring state consists of different phases of activity, namely, pre-flare, flare \& post-flare regions. $\gamma$-ray spectral analysis of these different activity phases has been performed and the best fit model for its spectra is found to be a Log-parabola model. We have also studied the correlation of simultaneous $\gamma$-ray light curves with the optical \& radio counterparts in these flaring states and report the DCF with 95\% significance level. A large time delay is found between radio and gamma ray data for two flares, indicating two zones of emission. We have fitted the multi-wavelength data with a two zone leptonic model. In our two zone leptonic model the maximum required power in the jet is 9.64 $\times$ 10$^{46}$ erg sec$^{-1}$, which is lower than its Eddington luminosity $2.29\times 10^{47}$ erg sec$^{-1}$. 

\end{abstract}

\keywords{galaxies: active; gamma-rays: galaxies; individuals: 4C+28.07}


\section{Introduction} 

Active Galactic nuclei (AGN) have been classified in various categories depending upon the angle between the observer line of sight and the jet axis. Blazars belong to a class of AGN where one of the jets is tilted towards the observer within a few degrees ($\sim 14^\circ$).
Blazar jets consist of highly energetic charged particles travelling along the jet axis with high Lorentz factor, which make the jets highly relativistic. The emissions produced inside the jets are also boosted along the jet axis due to high Doppler boosting, which amplifies the total radiation and as a result these sources can outshine their host galaxies. Even after decades of research the cause of acceleration of charged particles inside jets has remained unclear. Many models have been proposed in the past which includes diffusive shock acceleration and magnetic re-connection (see the \citealt{2019ARA&A..57..467B}) to explain the particle acceleration and eventually production of broadband flares. These models have been applied to many sources but still the underlying physics of multi-wavelength variability and flares is not well understood. Previous studies such as radio opacity and core-shift effect on blazars under the assumption of equipartition (\citet{2011A&A...532A..38S}, \citet{2014MNRAS.437.3396K}, \citet{2019ApJ...870...28F}) suggest that the jets have a strong magnetic field, of the order of a few Gauss, which is responsible for the synchrotron emission observed in optical, ultraviolet, and soft X-ray band. Blazars got more attention after the launch of the Fermi Gamma-ray Space telescope, which is solely designed to observed these sources in high energy $\gamma$-rays. The $\gamma$-rays produced in the jets of blazars may originate from leptonic or hadronic models. In leptonic models the gamma rays are produced by the inverse Compton scattering of the soft or low energy photons by the high energetic leptons (\citet{1992ApJ...397L...5M}, \citet{1993ApJ...416..458D}, \citet{1994ApJ...421..153S}). The low energy photons could be produced by synchrotron emission of relativistic leptons or they could be external photons from outside the jet produced in the broad line region (BLR) or dusty torus (DT), or in some cases it could be direct disk photons. In hadronic models the hadrons emit most of the high energy photons in the observed SED, which they can do by proton synchrotron (\citet{2000NewA....5..377A}, \citet{2001APh....15..121M}) or proton-proton (\citet{PhysRevD.47.5270}, \citet{1997MNRAS.287L...9B}, \citet{PhysRevD.80.083008}, \citet{2010ApJ...724.1517B})  or proton-photon (\citet{1992A&A...253L..21M}, \citet{1999MNRAS.302..373B}), \citet{2002PhRvD..66l3003N}, \citet{2003ApJ...586...79A}) interactions. The advantage of hadronic model is that it can also explain the observation of astrophysical neutrinos. As an example, blazar TXS 0506+056 has been identified as a source of astrophysical neutrinos and high energy $\gamma$-rays \citep{2018Sci...361..147I}.   

Past studies also suggest that the emission from blazars could be very complex involving multiple zones of emission. Sometimes the emission from the same source can be explained by the leptonic as well as hadronic model as it has been shown in \citet{2016arXiv160305506P}. 
Both leptonic and hadronic models can be used for single or multiple  zones of emission.
4C+28.07 is a FSRQ (Flat Spectrum Radio Quasars) type and moderately variable extra-galactic source with RA: 2h37m52.406s, DEC: +28d48m8.990s \citep{2004AJ....127.3587F}, located at red-shift 1.206 \citep{2012ApJ...748...49S} frequently monitored by Fermi Large Area Telescope (Fermi-LAT) since 2008. 
It was observed in radio with Effelsberg 100 m and the IRAM 30 m telescope since January and June 2007. It reached a high state in radio in mid 2008, subsequently the flux was decreasing until 2010, when the source reached the minima in radio flux \citep{2011ATel.3674....1N}. 


The Energetic Gamma Ray Experiment Telescope (EGRET) detected this source for the first time in high energy band ($> 100$ MeV) with maximum flux of (0.31$\pm$0.12)$\times$10$^{-6}$ photons cm$^{-2}$ s$^{-1}$.
\citet{2011ATel.3670....1S} reported the high flux state of this source with flux of (1.4$\pm$0.3$)\times 10^6$ photons cm$^{-2} s^{-1}$ on 3rd October, 2011 in $\gamma$-ray bands. This was an increase by a factor of 14 over the average flux of this source observed by Fermi LAT during the first two years of Fermi mission.

Swift observations on Oct 6 and Oct 9 of 2011 revealed an elevated  level of X-ray and optical activity \citep{2011ATel.3676....1S}. 4C+28.07 was also detected with the Einstein Observatory Image Proportional Counter (IPC; 0.2-3.5 keV) at 1980. Though this detection in IPC may not be reliable. This source was observed on Dec 23 in 2013 when it was flaring in NIR. The observations were carried out by 2.1m telescope of the Guillermo Haro Observatory \citep{2014ATel.5711....1C}. This source has been also continuously monitored in 15 GHz radio band by MOJVAE survey \citep{2019ApJ...874...43L} since 1994. The maximum flux density at 15 GHz and maximum jet speed were obtained as 5.11 Jy and 409$\pm$24 $\mu$as yr$^{-1}$ between 1994 to 2019. The Fermi-GST AGN Multi-frequency Monitoring Alliance (F-GAMMA) programme (\citet{2010arXiv1006.5610A}, \citet{2014MNRAS.441.1899F}, \citet{2015A&A...575A..55A}) provided the multi-frequency radio light curve (2.64 GHz to 43 GHz) of 4C+28.07 between 2007 and 2015, which was released in the second part of the F-GAMMA data-set \citep{2019A&A...626A..60A}. Candidate Gamma-Ray Blazar Survey Source Catalog (CGRABS) recorded R band magnitude of 16.99 \citep{2008ApJS..175...97H}.

In this paper we have analysed the Fermi LAT data collected over a period of 12 years from this source, identified the flares in its long term light curve and subsequently modelled them with multi-wavelength data.

\section{DATA ANALYSIS}

\subsection{FERMI-LAT ANALYSIS}
 Fermi-Large Area Telescope (LAT) is a pair conversion imaging  $\gamma$-ray telescope, which covers an energy range from 20 MeV to $>$300 GeV with effective area of
 $>8000$ cm$^2$ \citep{2009ApJ...697.1071A} and scans the entire sky in survey mode with a time period of $\sim$ 3.2 hours. Further detailed characteristics of LAT-instrument is provided on the Fermi Webpage \footnote{https://fermi.gsfc.nasa.gov/ssc/data/analysis/software/}. We have extracted the data of FSRQ 4C+28.07 from FSSC's  website data server \footnote{https://fermi.gsfc.nasa.gov/cgi-bin/ssc/LAT/LATDataQuery.cgi} over the period of $\sim$ 12 years (August, 2008 - May, 2020) and analyzed it with the help of Fermi science tool software package version- 1.0.10, which includes galactic diffuse emission model (\texttt{gll\_iem\_v07.fits}) and extra galactic isotropic diffuse emission model (\texttt{iso\_P8R3\_SOURCE\_V2\_v1.txt}). 
 The ``unbinned Likelihood analysis" method is used to analyze the Fermi-LAT Pass8 data with appropriate  selections and cuts. The photon-like events are classified as 
 ‘evclass=128, evtype=3’ with energies range from 100 MeV to 300 GeV. We have extracted the photons from a radius (Region of interest or ROI) of 10$^\circ$ 
 around the source and used maximum zenith angle value of 90$^\circ$, which is the standard value provided by the LAT-instrument team, in order to avoid the $\gamma$-ray detection from the earth's limb.  Filter expression ``\texttt{DATA\_QUAL}$>$0 \&\& \texttt{LAT\_CONFIG}==1 \&\& \texttt{ANGSEP}(\texttt{{}RA\_SUN},\texttt{DEC\_SUN},39.4684,28.8025)$>$15" is implemented to select the good time interval data and to avoid time bins when the Sun could be close to the target (less than 15$^\circ$), which is recommended by the LAT team. The moon is brighter in gamma-ray above 31 MeV\footnote{https://www.nasa.gov/feature/goddard/2019/moon-glows-brighter-than-sun-in-images-from-nasas-fermi}, we have also checked the effect of moon on our results between 100 MeV to 300 GeV and did not find any significant contamination. We have further followed the same procedure as described in \cite{2020ApJS..248....8D} with the latest 4FGL catalog  \citep{2015ApJS..218...23A}.


\subsection{SWIFT-XRT/UVOT}
 We have also analyzed the archival data from the X-ray Telescope (XRT) and Ultraviolet-Optical Telescope (UVOT) of The Neil Gehrels Swift Observatory, which is a space based telescope observing galactic and extra-galactic sources in X-ray, UV \& Optical wavelength. The data have been retrieved from HEASARC webpage\footnote{https://heasarc.gsfc.nasa.gov/cgi-bin/W3Browse/swift.pl} during the time span of same as $\gamma$-ray period and total 12 observations were made. A task called `$xrtpipeline$' (version of 0.13.2.) was used to process the XRT-data (\citet{2005SSRv..120..165B}, 0.3-10 keV) files for each observation-id. The calibration files with version of 20160609 and standard screening criteria have been implemented in this process. We have 
 chosen a circular radius of 25 arc seconds and background region as annular ring around the source to analyze these data. The tools `$xselect$' and `$xrtmkarf$' have been used to extract the light curve \& and to  create the ancillary response file respectively. We have grouped the spectra of 20 counts per bin by `$grppha$' and after that these grouped spectra have been modelled in XSPEC (version of 12.11.0) with the model - `$tbabs*log parabola$'. We use the neutral hydrogen column density of $n_H = 7.75\times10^{20} cm^{-2}$, provided in HEASARC webpage \footnote{https://heasarc.gsfc.nasa.gov/cgi-bin/Tools/w3nh/w3nh.pl}.

Simultaneous observation was also made for this source by the Swift Ultraviolet/Optical telescope (UVOT, \citet{2005SSRv..120...95R}) in all six filters: U, V, B, W1, M2, 
\& W2. The source \& background region have been extracted from circular region of 5 arc seconds and an annular region around the source respectively. The task `$uvotsource$' is used to extract the source magnitudes and `$uvotimsum$' to sum up the more than one observation in particular filter. These magnitudes are then corrected for galactic extinction 
\citep{2011ApJ...737..103S} and subsequently converted into flux by using the zero points \citep{2011AIPC.1358..373B} and proper conversion factors \citep{2016MNRAS.461.3047L}.

\subsection{OPTICAL DATA}
We have made use of publicly available archival data of Catatlina surveys\footnote{http://nesssi.cacr.caltech.edu/DataRelease/} and Steward Optical Observatory, Arizona \footnote{http://james.as.arizona.edu/ psmith/Fermi/} \citep{2009arXiv0912.3621S}, which are a part of {\it Fermi} monitoring program. Catalina provides V-band photometric data. Steward observatory provides V, R - band photometric and polarimetric data (Degree of polarization \& Position angle).

We have collected all the available data of this source 4C+28.07 during MJD 54406 - 56590. There is no V \& R - band light curve data available in the Steward observatory website.

\subsection{OVRO}
We have also collected (MJD 54473 - 59109) publicly available data from the 40 meter ground based telescope of Owens Valley Radio Observatory \footnote{http://www.astro.caltech.edu/ovroblazars/}(OVRO; \citet{2011ApJS..194...29R}), which regularly observe several blazars in 15-GHz radio band as a part of Fermi-monitoring program.

\section{Flaring State In $\gamma$-ray Light curves}
\begin{figure*}[h] 
\centering
\includegraphics[height=2.6in,width=5.3in]{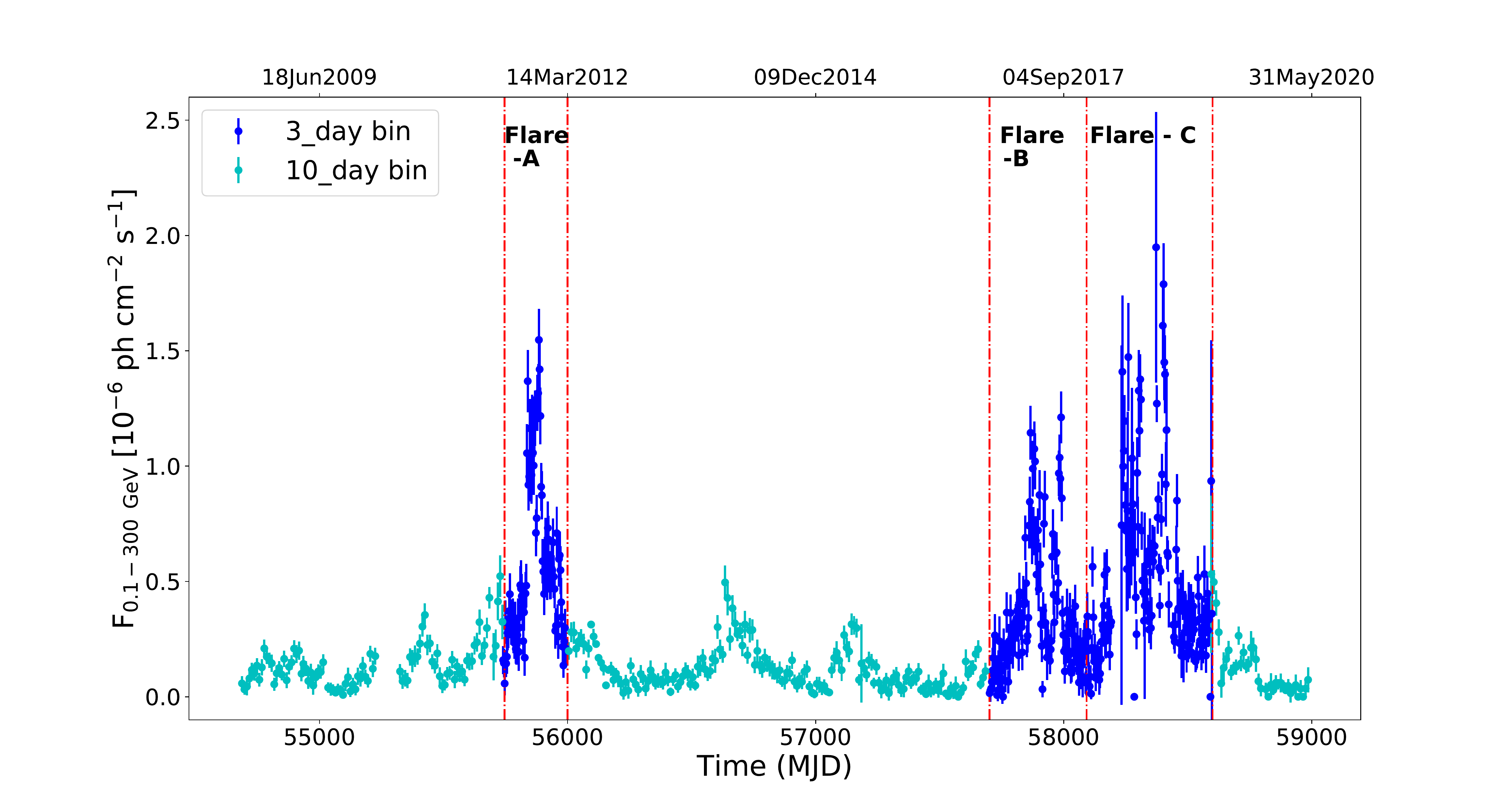}
\caption{Fermi-LAT light curve of 4C+28.07 (MJD 54682 - 59000). Three major flaring states have been identified, which are shown by broken red line. Cyan and blue color represent the data in 10-day \& 3-day binning respectively.}
\label{fig:1}
\end{figure*}

We have analyzed the Fermi-LAT data of 4C+28.07 during MJD 54682 - 59000 ($\sim$ 12 yr) in 10 days binning and found three distinctive flaring states: Flare-A, Flare-B \& Flare-C. These flaring states have been further analyzed in 3 day binning to study the different activity phases extensively. Figure-\ref{fig:1} shows the 12 year light curve of this source with flaring and quiescent state in different color.

We have fitted each Flare phase with the following function \citep{2010ApJ...722..520A} to compute the rising ($T_r$) \& decay time ($T_d$),

\begin{equation} \label{eq:1}
F(t)=2F_0\Big[\exp \Big(\frac{t_0-t}{T_r}\Big)+\exp \Big(\frac{t-t_0}{T_d}\Big)\Big]^{-1} 
\end{equation}

Here, $F_0$ \& $t_0$ are the photon flux value and time at the given peak respectively. 

We have used Bayesian block representation to detect the different states of activity, which is based on \citet{2013ApJ...764..167S} algorithm and used before by several authors (e.g. \citet{2019ApJ...877...39M}, \citet{2020A&A...642A.129S}).
In this paper, all the reported $\gamma$-ray photon fluxes are in the unit of $10^{-6}$ ph cm$^{2}$ s$^{-1}$. 
\subsection{FLARE-A}
\begin{figure*}[h] 
\centering

\includegraphics[height=2.6in,width=5.3in]{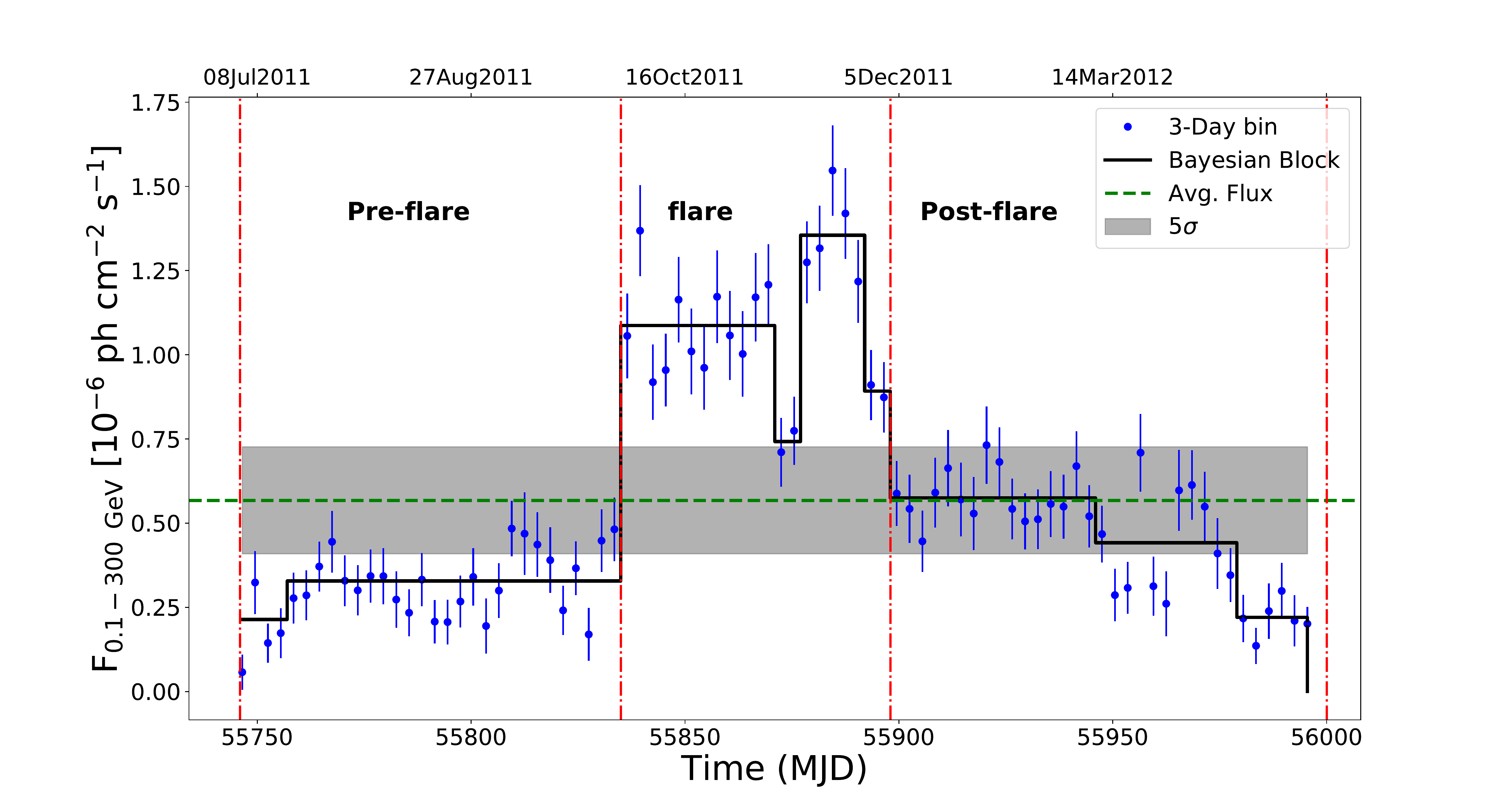}
\caption{Three day binning light curve of Flare-A. Time duration of all the different periods of activities, which are shown by broken red line :  MJD 55746-55835 (Pre-flare), MJD 55835-55898 (Flare) and MJD 55898-56000 (Post-flare).}
\label{fig:2}

\includegraphics[height=2.6in,width=5.3in]{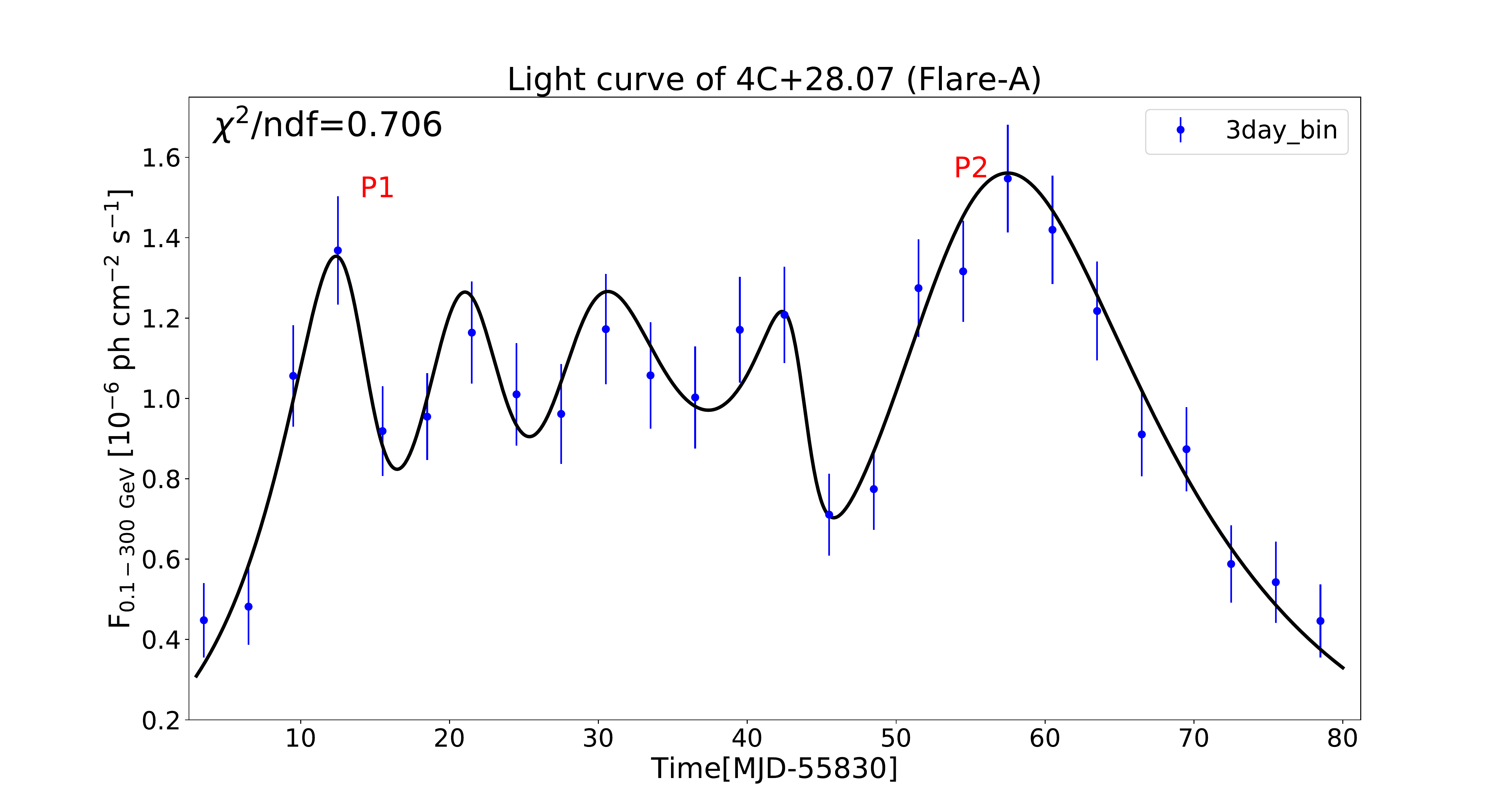}
\caption{Fitted light curve of flare phase (Flare-A) with time span of 71 days (MJD 55835-55906).}
\label{fig:3}

\includegraphics[height=2.6in,width=5.3in]{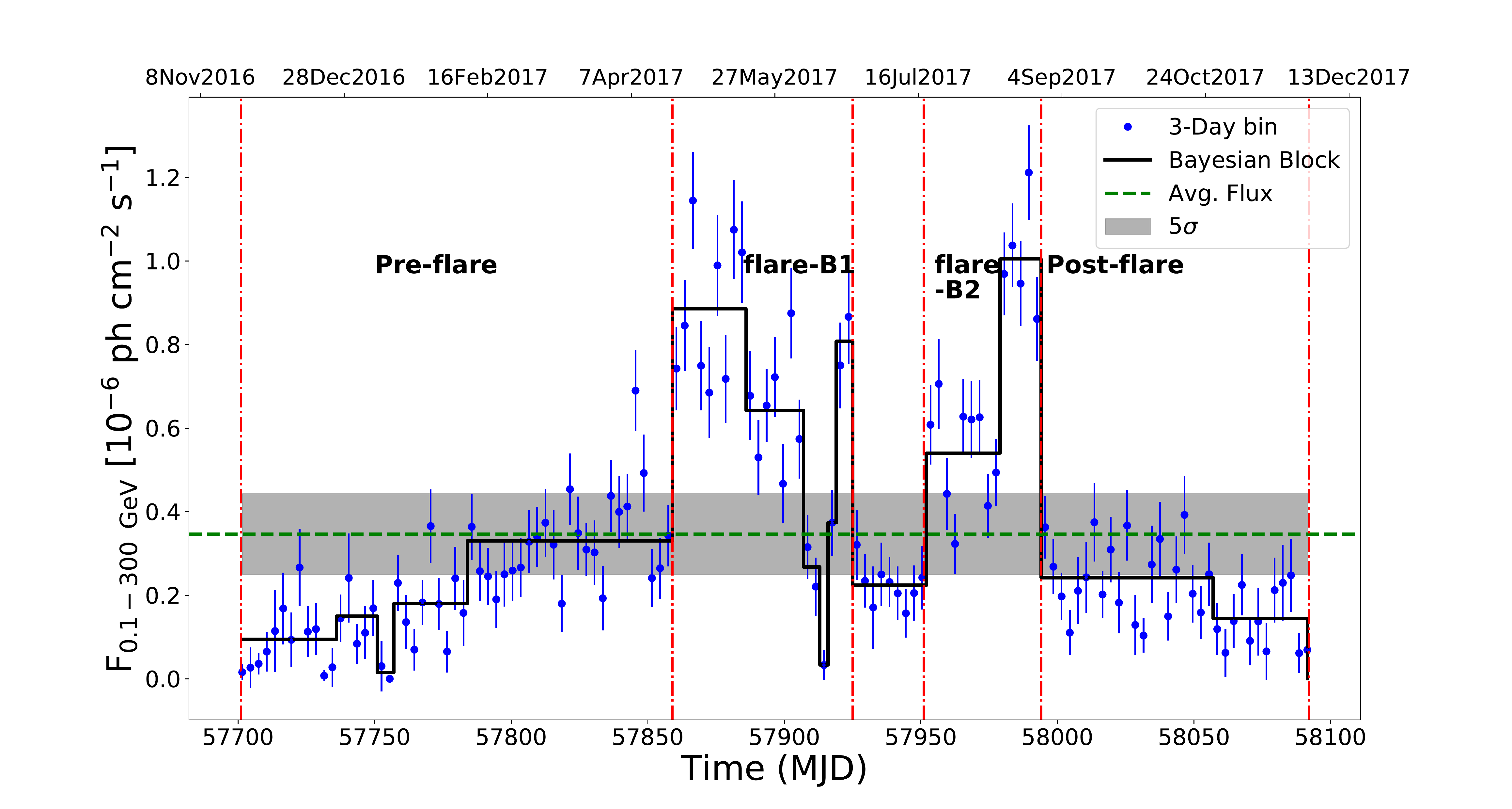}
\caption{Three day binning light curve of Flare-B. Time duration of all the different periods of activities, which are shown by broken red line:  MJD 57701-57859 (Pre-flare), MJD 57859-57925 (Flare-B1), MJD 57952-57994 (Flare-B2) and MJD 57994-58092 (Post-flare).}
\label{fig:4}

\end{figure*}

A 3-day binning light curve of Flare-A is shown in Figure-\ref{fig:2}, which has three different phases, defined as Pre-flare, Flare \& Post-flare.

Pre-flare phase lasted for $\sim$ 80 days (MJD 55746-55835) with average flux of 0.30$\pm$0.08. Subsequently, the source entered the flare phase with significant variation in flux, which lasted for $\sim$ 71 days (MJD 55835 - 55898). Figure-\ref{fig:3} shows the fitted light curve of this Flare phase. We have identified two major peaks (P1 \& P2) with fluxes of 1.37$\pm$0.13, 1.55$\pm$0.13 at MJD 55839.5 \& 55884.5 respectively. The fitted parameters of the light curve have been described in Table-\ref{tab:1}.

\subsection{FLARE-B}
Similar to Flare-A, a 3 day binning analysis has also been carried out for Flare-B and four different phases have been observed. Figure-\ref{fig:4} shows these phases, named as Pre-flare, Flare-B1, Flare-B2 \& Post-flare.

\begin{figure*}[h]
\centering

\includegraphics[height=2.6in,width=5.3in]{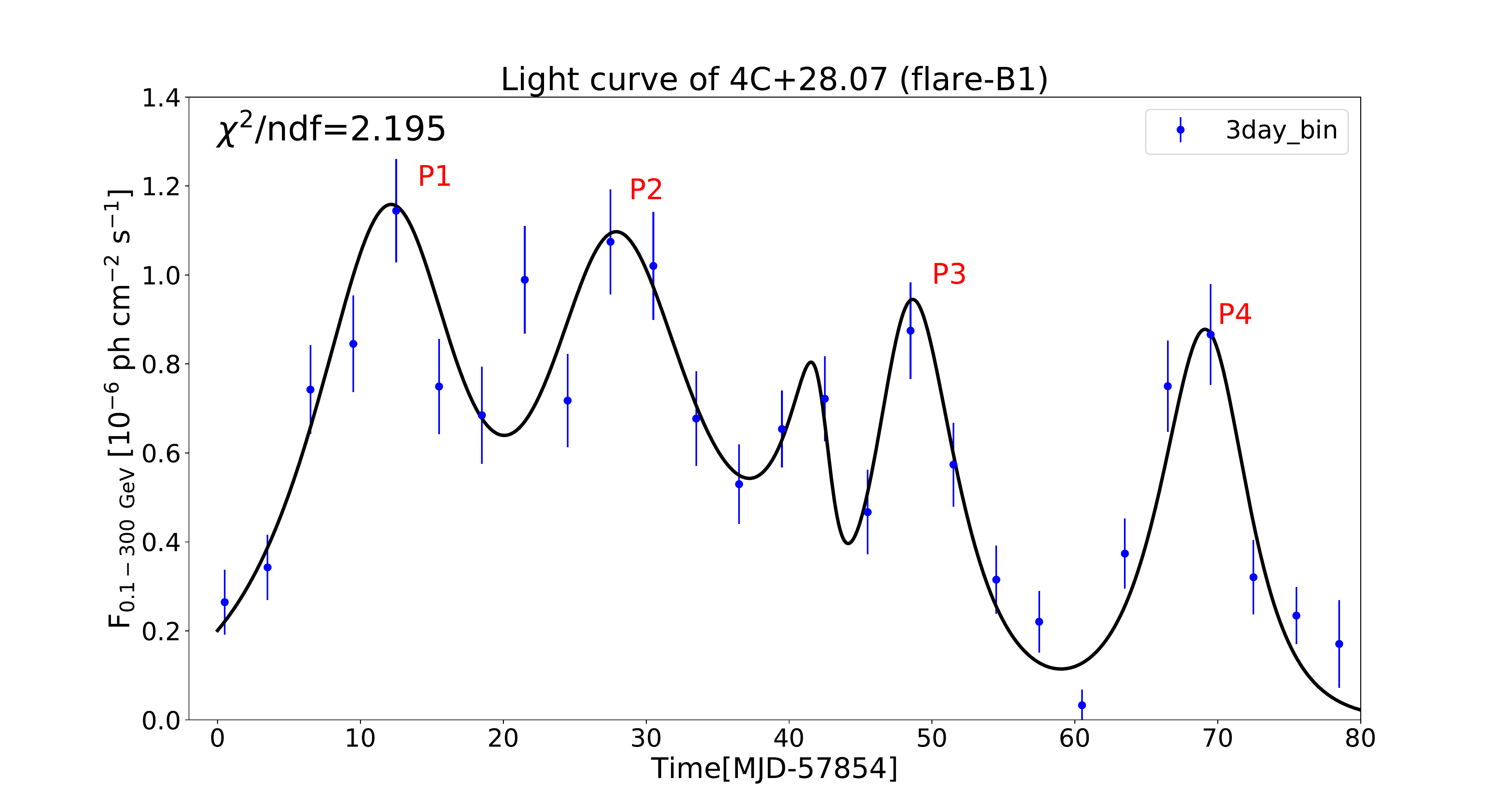}
\caption{Fitted light curve of flare-B1 phase (Flare-B) with time span of 66 days (MJD 57859-57925).}
\label{fig:5}

\includegraphics[height=2.8in,width=5.3in]{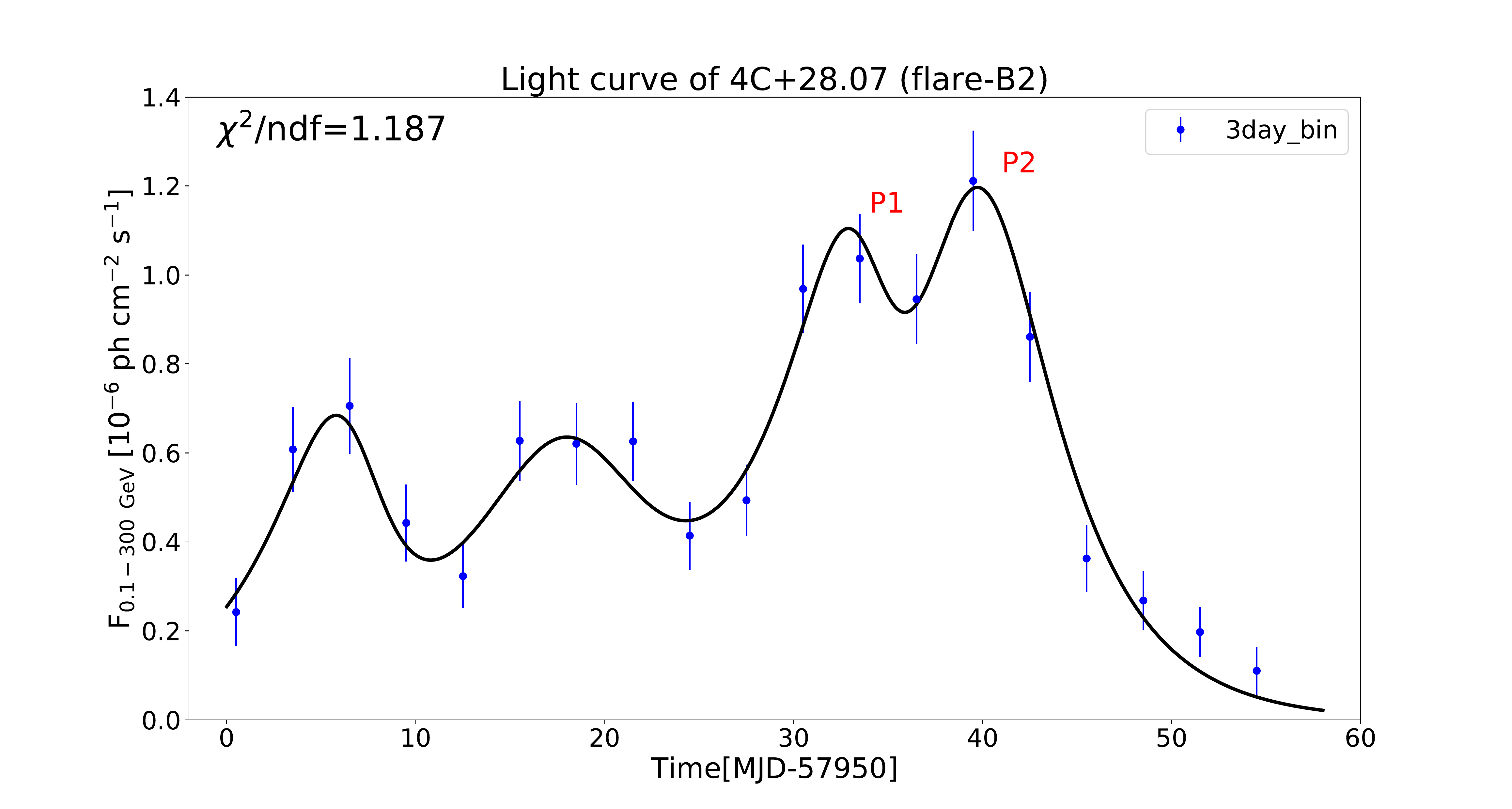}
\caption{Fitted light curve of flare-B2 phase (Flare-B) with time span of 42 days (MJD 57925-57994).}
\label{fig:6}

\includegraphics[height=2.6in,width=5.3in]{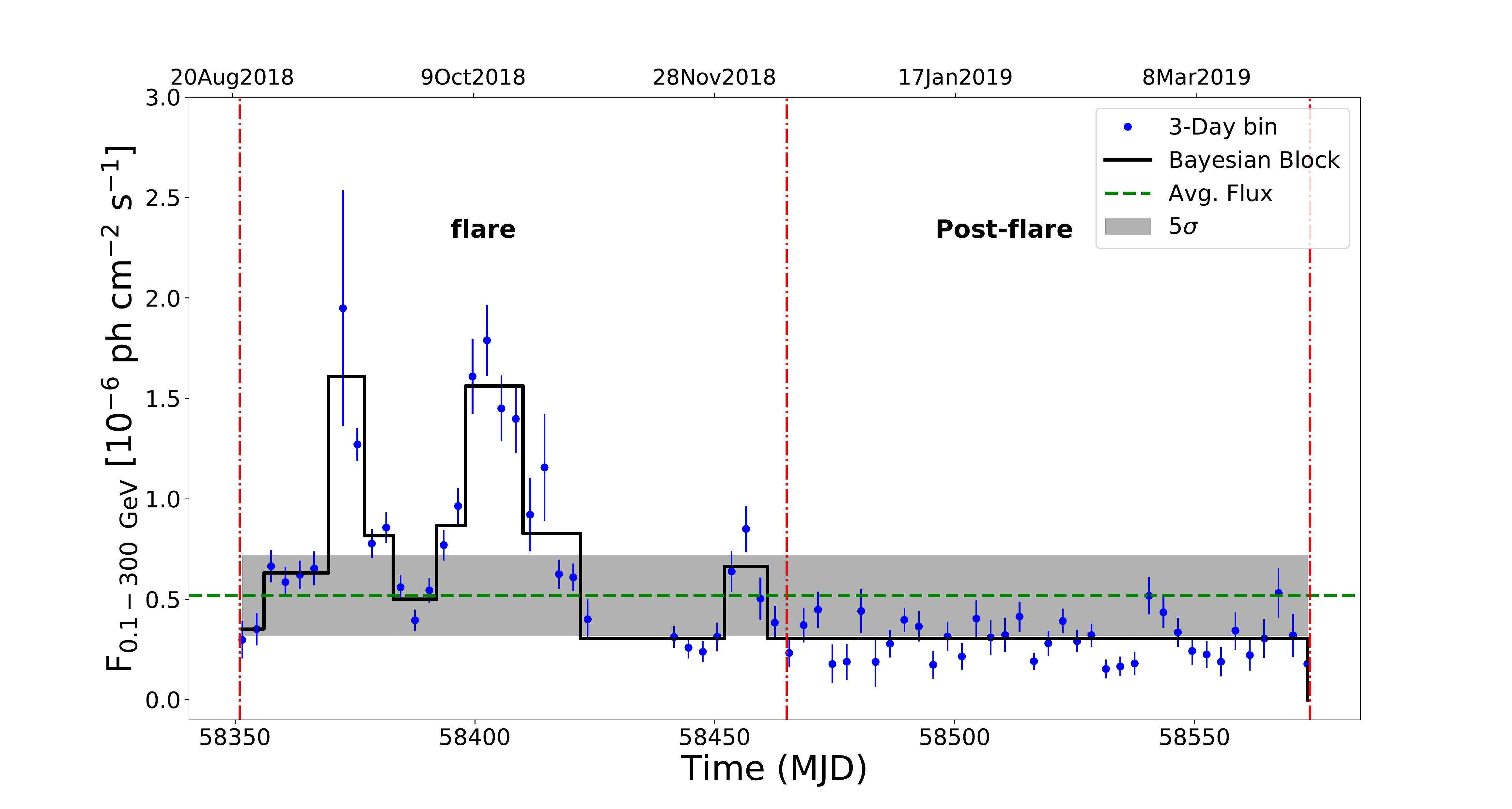}
\caption{Three day binning light curve of Flare-C. Time duration of all the different periods of activities, which are shown by broken red line: MJD 58355-58422 (Flare) and MJD 58421-58574 (Post-flare).}
\label{fig:7}

\end{figure*}

The Pre-flare phase of Flare-B has very small variation in flux with time span of $\sim$ 132 days (MJD 57701 - 57859). Fitted light curve of Flare-B1 \& Flare-B2 have been illustrated in Figure-\ref{fig:5} \& Figure-\ref{fig:6} respectively. Flare-B1 consists of four major peaks (P1, P2, P3 \& P4) at MJD 57866.5, 57881.5, 57902.5 \& 57923.5 with fluxes of 1.14$\pm$0.12, 1.07$\pm$0.12, 0.84$\pm$0.11 \& 0.87$\pm$0.11 respectively. After flare-B1  the source entered a low flux state which lasted around $\sim$ 27 days and again rose to a high state, defined as flare-B2 (MJD 57952-57994). It has two distinctive peaks at MJD 57983.5, 57989.5. Post-flare phase (MJD 57994-58092) has also been shown in Figure-\ref{fig:4}. The values of the fitted parameters have been elucidated in Table-\ref{tab:2}.

\subsection{FLARE-C}

Flare-C has only two phases: flare \& Post-flare (shown in Figure-\ref{fig:7}). Due to irregularity and large error in the light curve data, we are unable to show the Pre-flare phase for this flaring state.

\begin{figure*}[h]
\centering
\includegraphics[height=2.8in,width=5.3in]{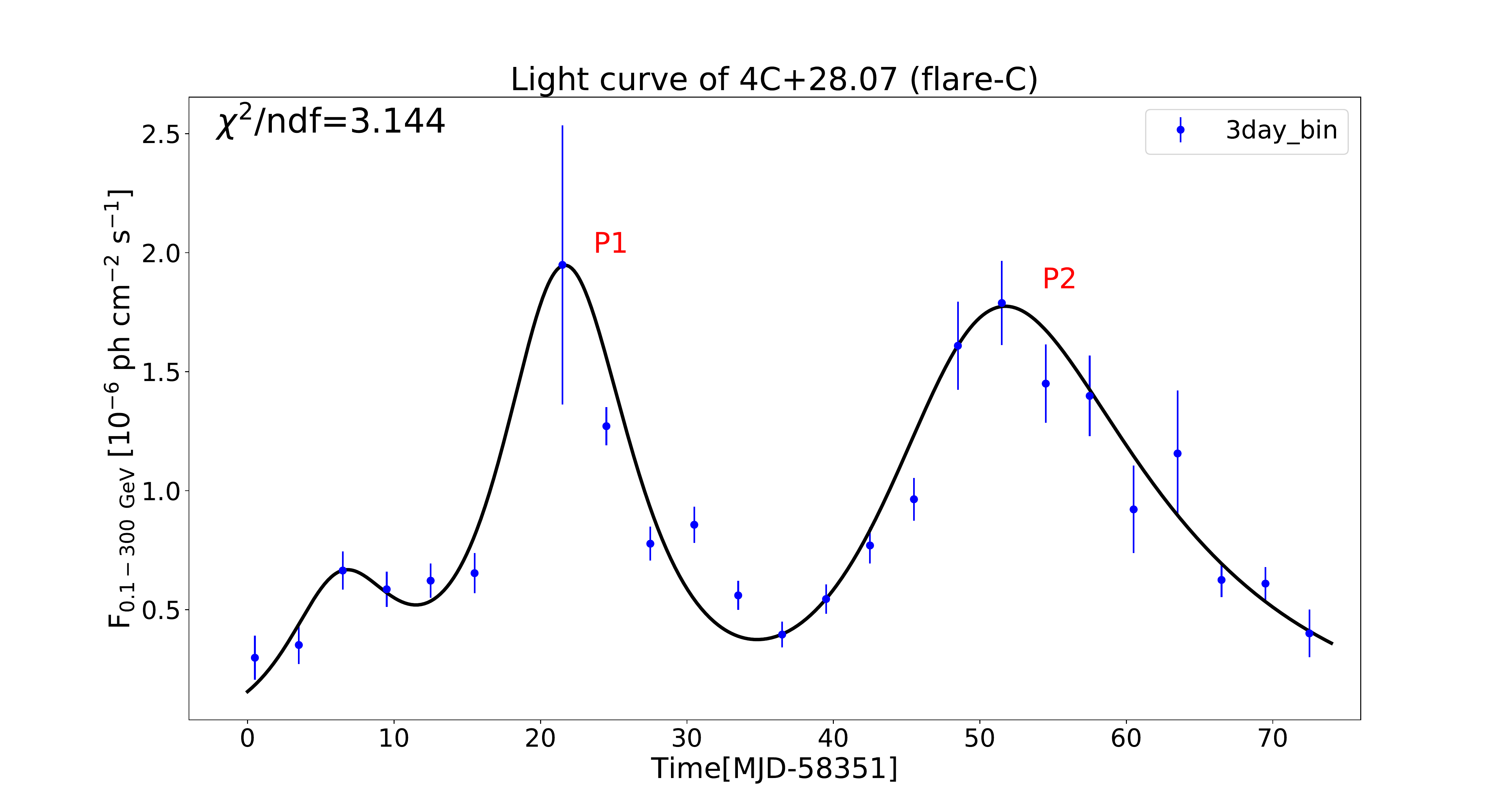}
\caption{Fitted light curve of flare phase (Flare-C) with time span of 66 days (MJD 58355-58421).}
\label{fig:8}

\includegraphics[height=1.90in,width=2.6in]{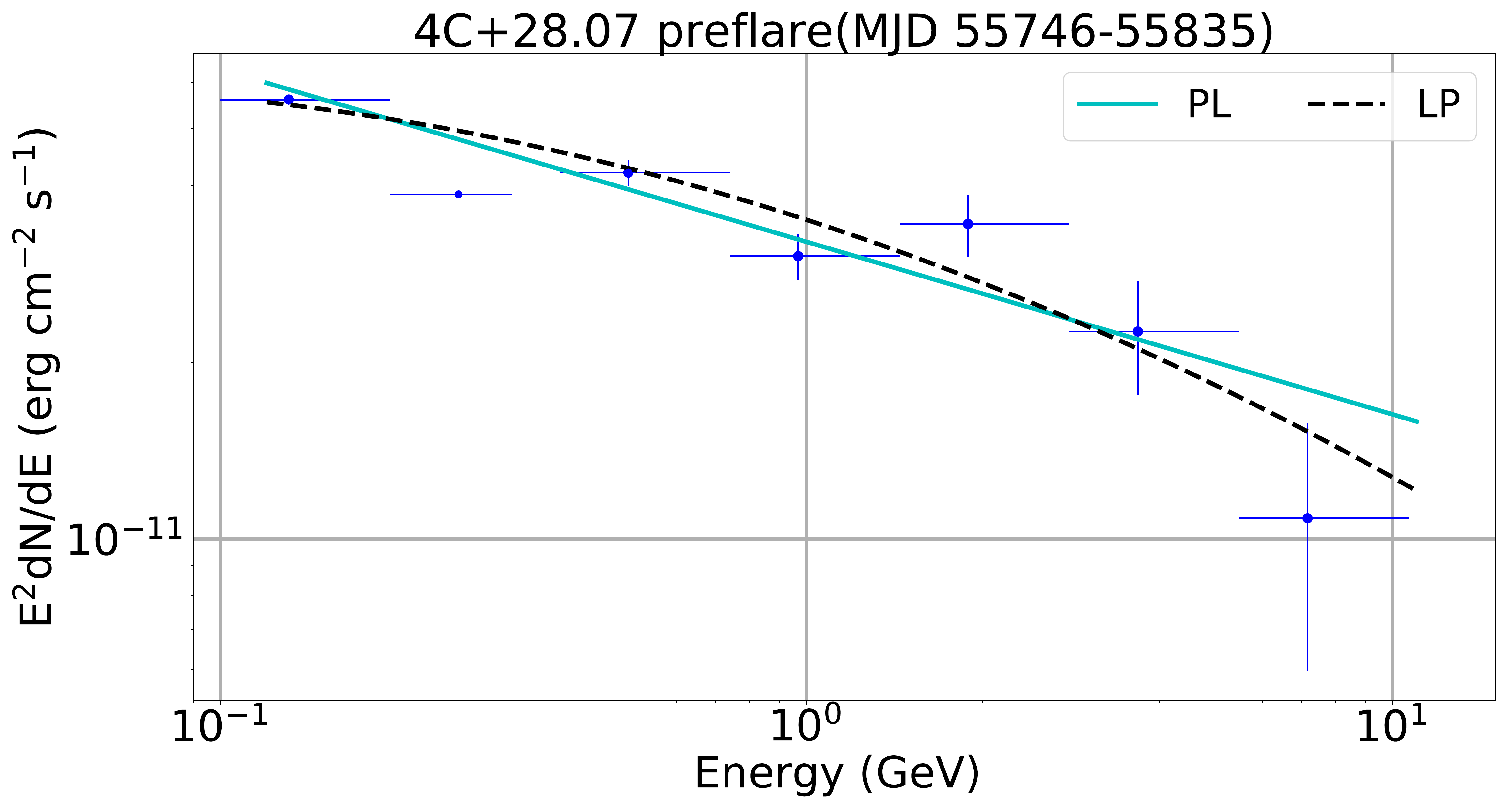}
\includegraphics[height=1.90in,width=2.6in]{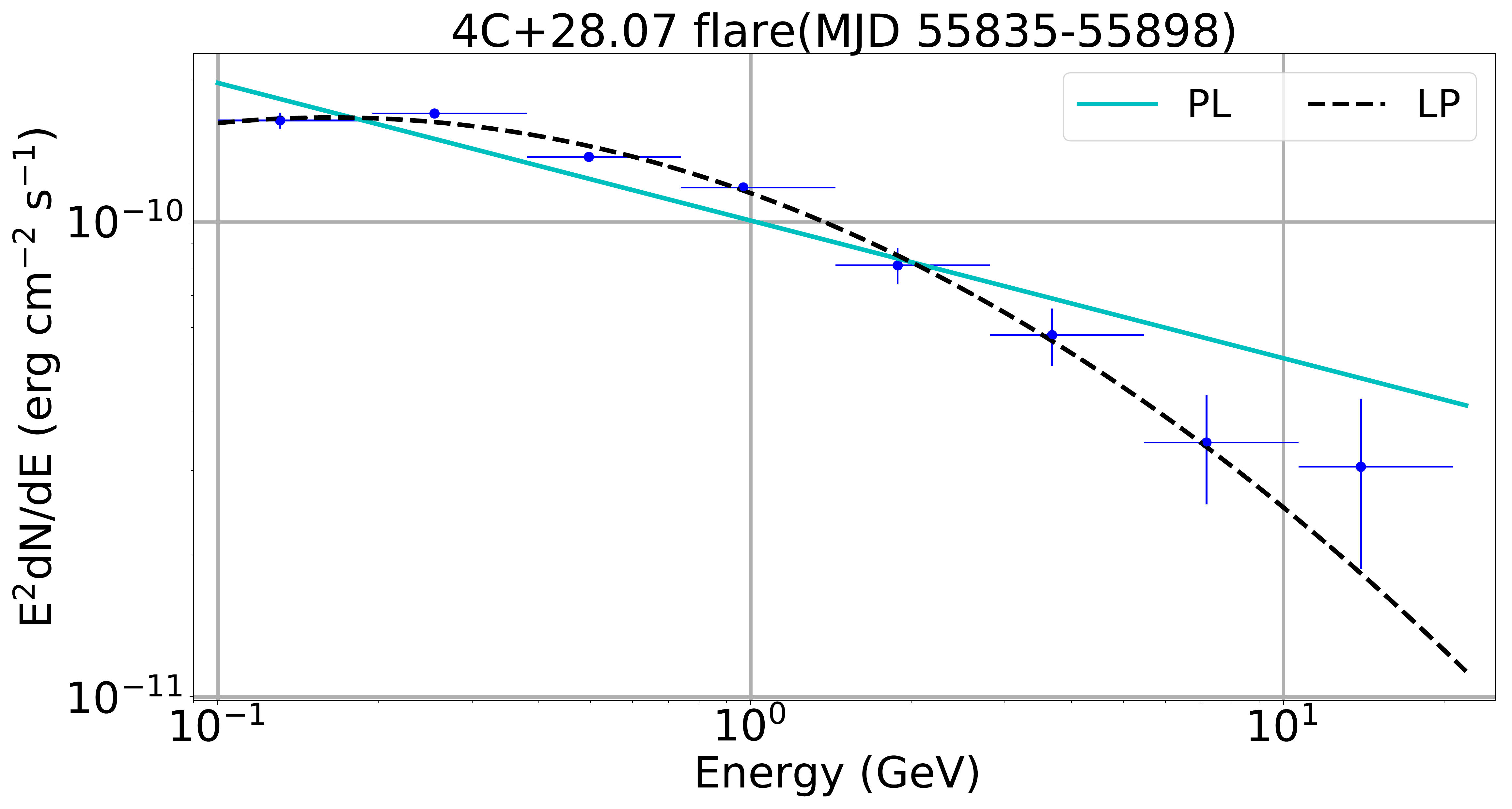}
\includegraphics[height=1.90in,width=2.6in]{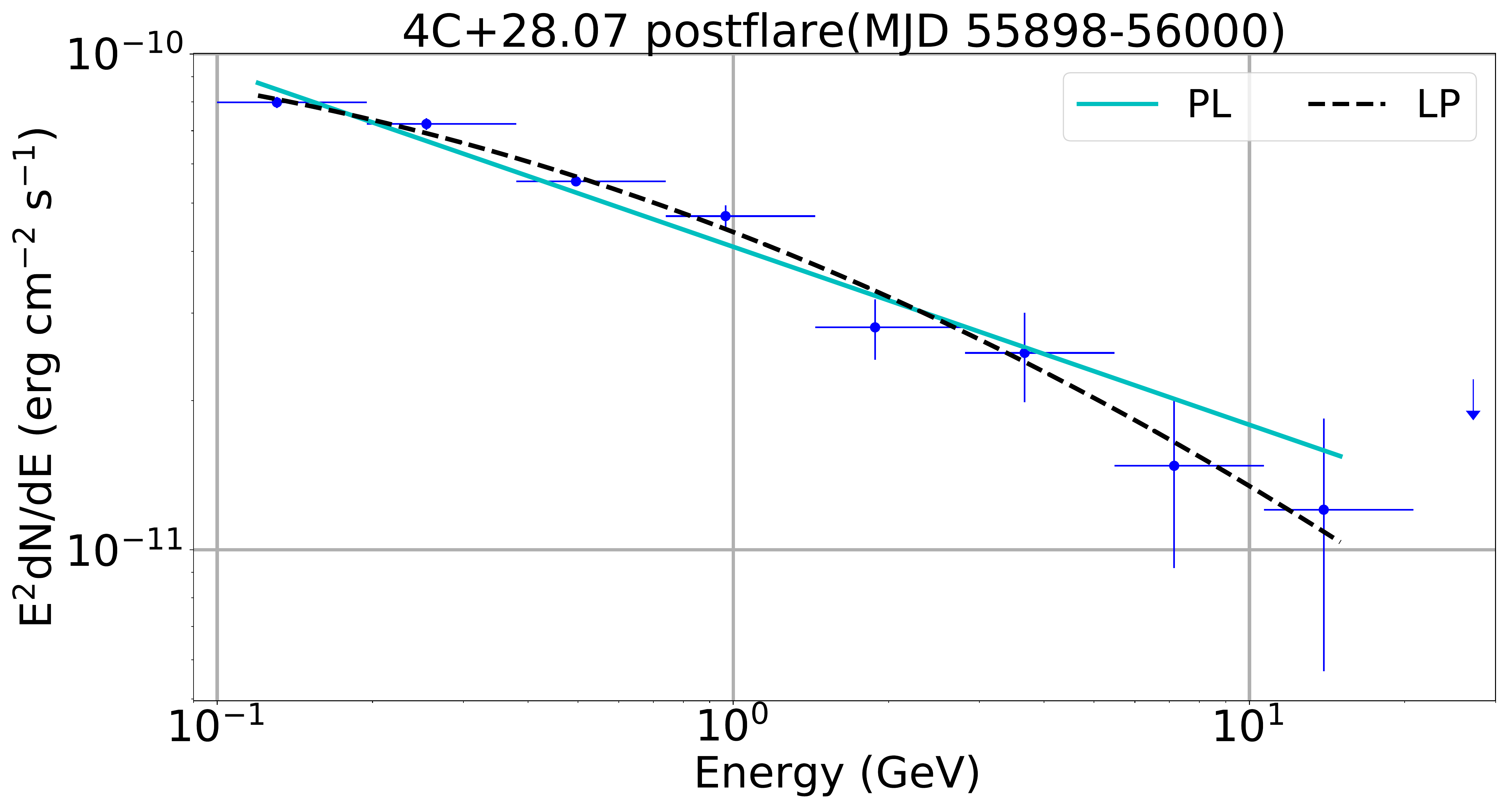}
\caption{$\gamma$-ray SED of different periods of activity (Pre-flare, flare, Post-flare) of Flare-A. PL, LP describe the Powerlaw, Logparabola model respectively, which are fitted to data points.}
\label{fig:9}

\end{figure*}

Fitted light curve of Flare phase has been shown in Figure-\ref{fig:8}, which has two major peaks. One peak (P1) was observed at MJD 58372.5 with flux of 1.95$\pm$0.57. This is the highest observed flux value over the 12 years period analyzed here. Another peak (P2) was found at MJD 58402.5. After this the source came back to low flux level and entered into Post-flare phase (MJD 58422-58574) with average flux of 0.33$\pm$0.08. The values of the fitted parameters are given in Table-\ref{tab:3}.


\section{$\gamma$-ray SED}
We have performed spectral analysis in $\gamma$-ray band for different phases (e.g. Pre-flare, Flare, Post-flare) of the flaring states and fitted these SED (Spectral Energy Distribution) with two different models \citep{2010ApJ...710.1271A}:

\begin{itemize}

\item A simple powerlaw model,

\begin{equation}
   \frac{dN}{dE} = N_{PL}\Big(\frac{E}{E_0}\Big)^{-\Gamma_{PL}}
\end{equation}

In our study, we have fixed the value of scaling factor ($E_0$) at 100 Mev. $N_{PL}$ \& $\Gamma_{PL}$ are the normalization and spectral index of the model respectively.

\item A log-parabola model, given by,

\begin{equation}
   \frac{dN}{dE} = N_{LP}\Big(\frac{E}{E_s}\Big)^{-(\alpha+\beta\log(E/E_s))}
\end{equation}

where, $L_0$, $\alpha$, $\beta$ are the normalization, photon and curvature index respectively. We have kept a fixed value of scaling factor ($E_s$) near the low energy part of the $\gamma$-ray SED (300 MeV).




\end{itemize}

We have fitted each SED of different phases with the above mentioned models: PL (Power-law) and LP (Log parabola) 

Figure-\ref{fig:9} shows SEDs (Pre-flare, Flare, Post-flare) of Flare-A. Spectral index (`$\Gamma_{PL}$' in PL model) is constant during the change of Pre-flare to Flare phase. However $\Gamma_{PL}$ changes from 2.29$\pm$0.04 to 2.36$\pm$0.03 as the source transited from Flare to Post-flare phase, which is a sign of spectral hardening behaviour. The fitted Parameters of different models have been elucidated in Table-\ref{tab:4}.

\begin{figure*}[h!]
\centering

\includegraphics[height=1.90in,width=2.6in]{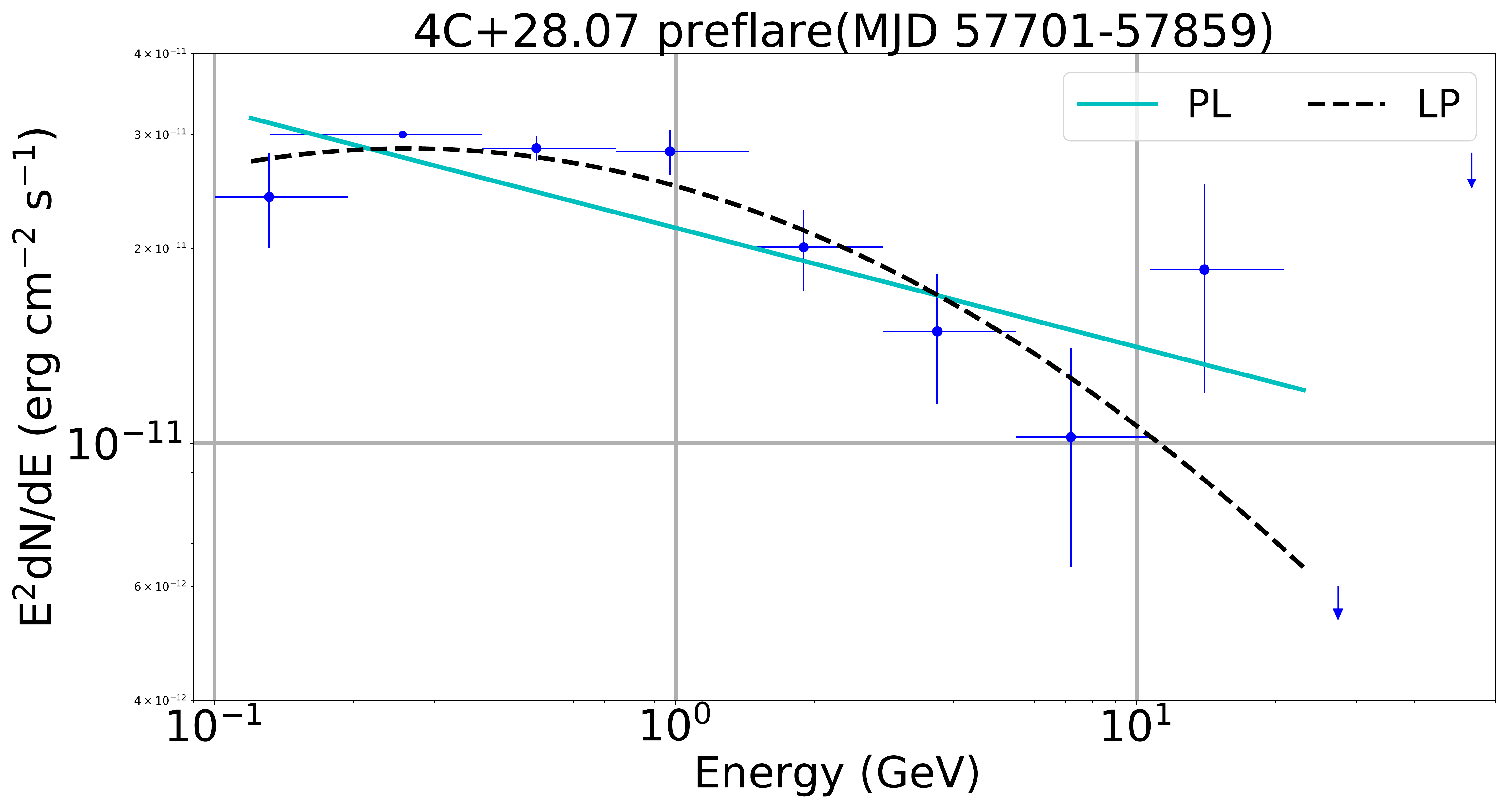}
\includegraphics[height=1.90in,width=2.6in]{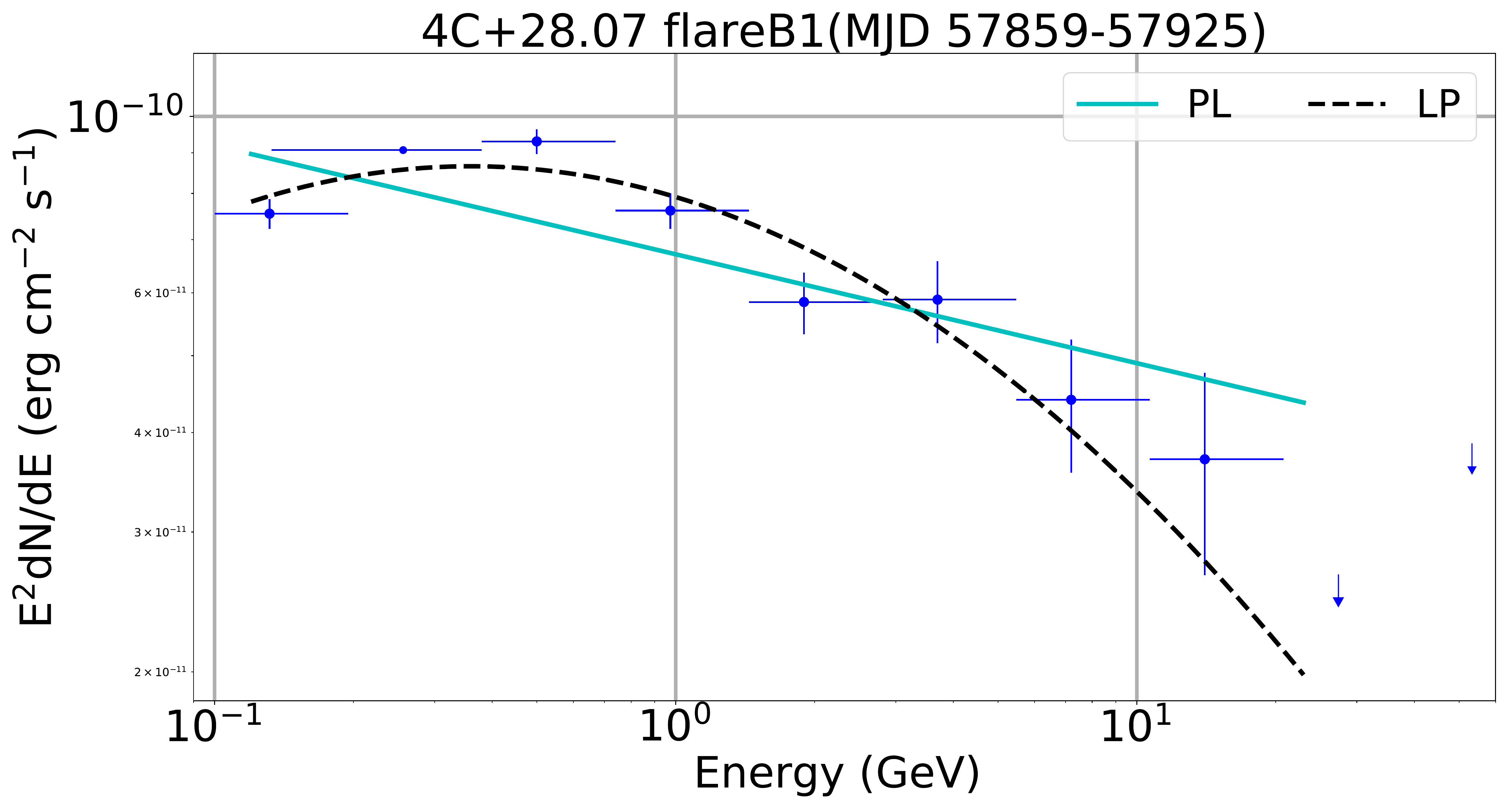}
\includegraphics[height=1.90in,width=2.6in]{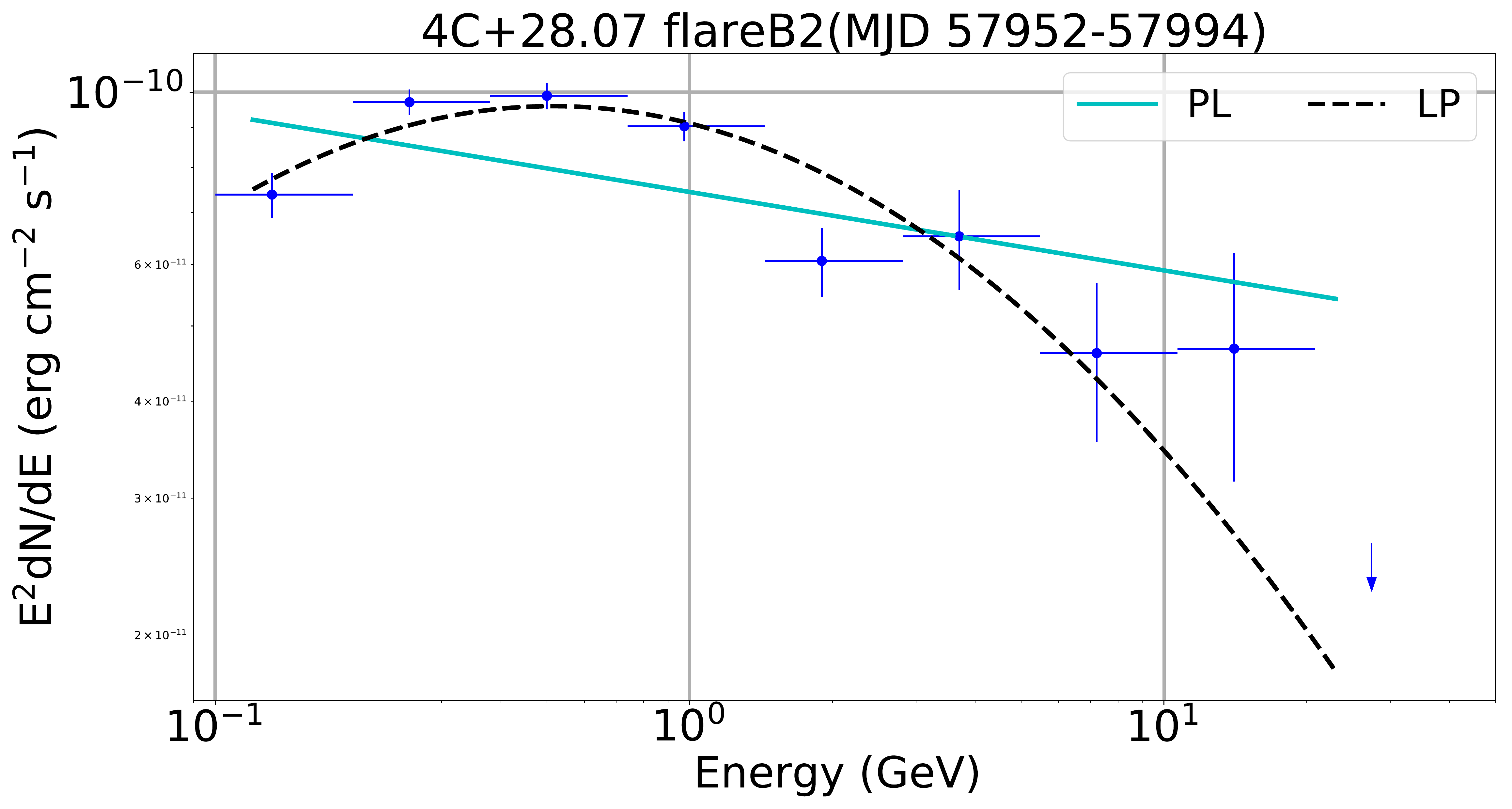}
\includegraphics[height=1.90in,width=2.6in]{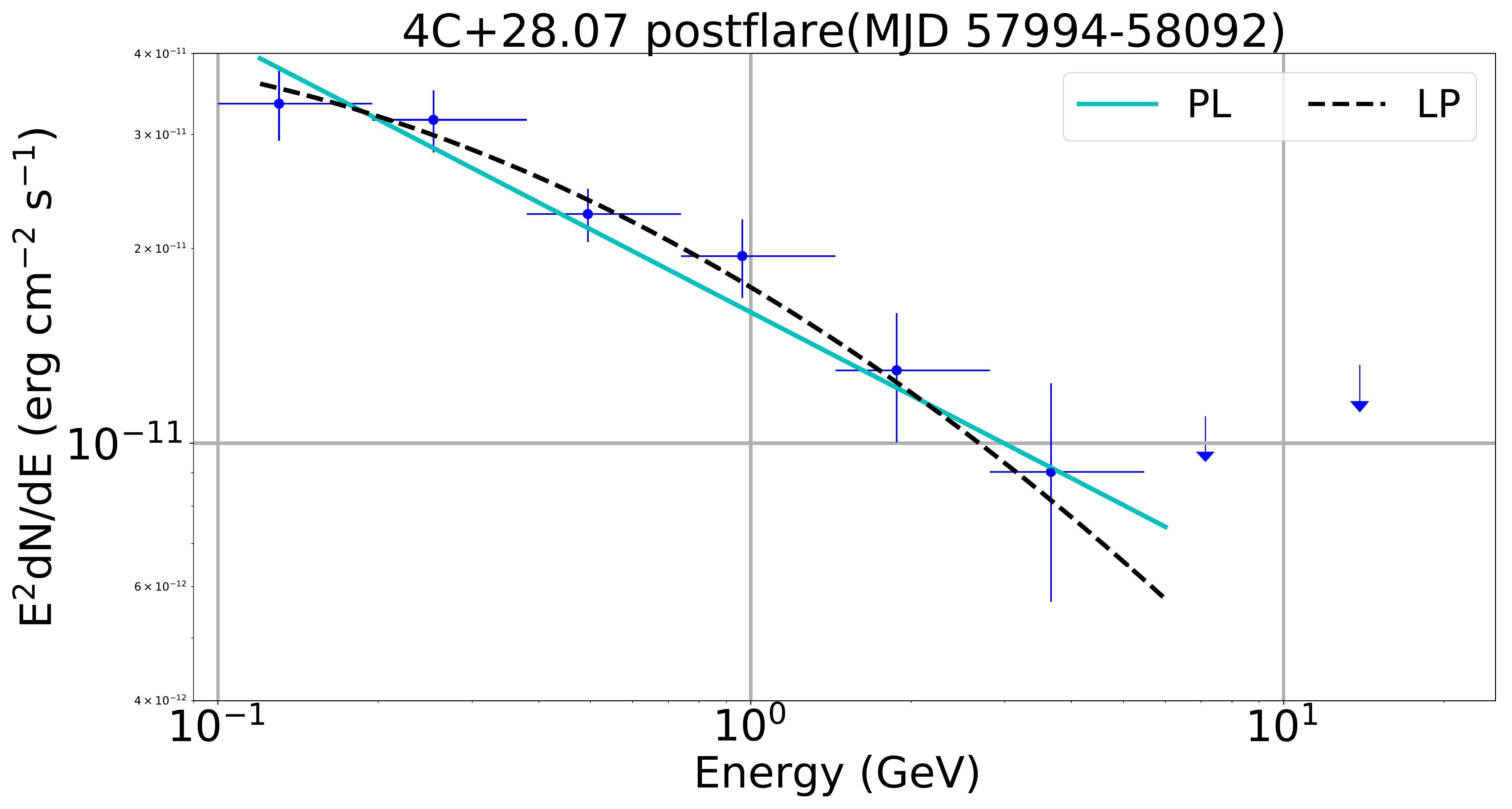}
\caption{$\gamma$-ray SED of different periods of activity (Pre-flare, flare-B1, flare-B2, Post-flare) of Flare-B. PL, LP describe the Powerlaw, Logparabola model respectively, which are fitted to data points.}
\label{fig:10}
\end{figure*}

All the four phases of Flare-B show significant spectral hardening with increasing flux ($\Gamma_{PL}$ $\propto$ $F_0^{-1}$) value, which are illustrated in Figure-\ref{fig:10}. The detailed description of model parameters have been shown in Table-\ref{tab:5}.

\begin{figure*}[h!]
\centering
\includegraphics[height=1.90in,width=2.6in]{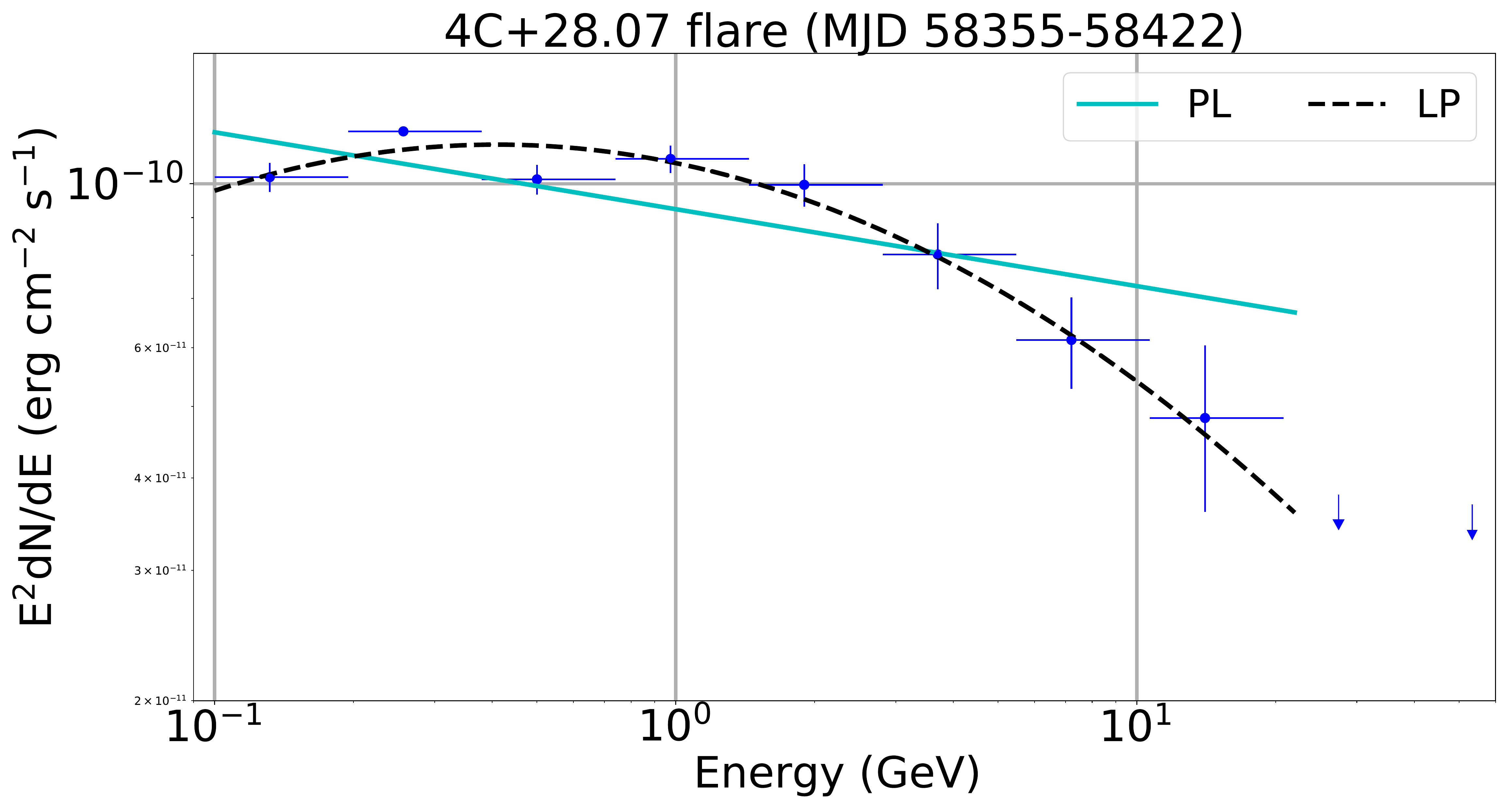}
\includegraphics[height=1.90in,width=2.6in]{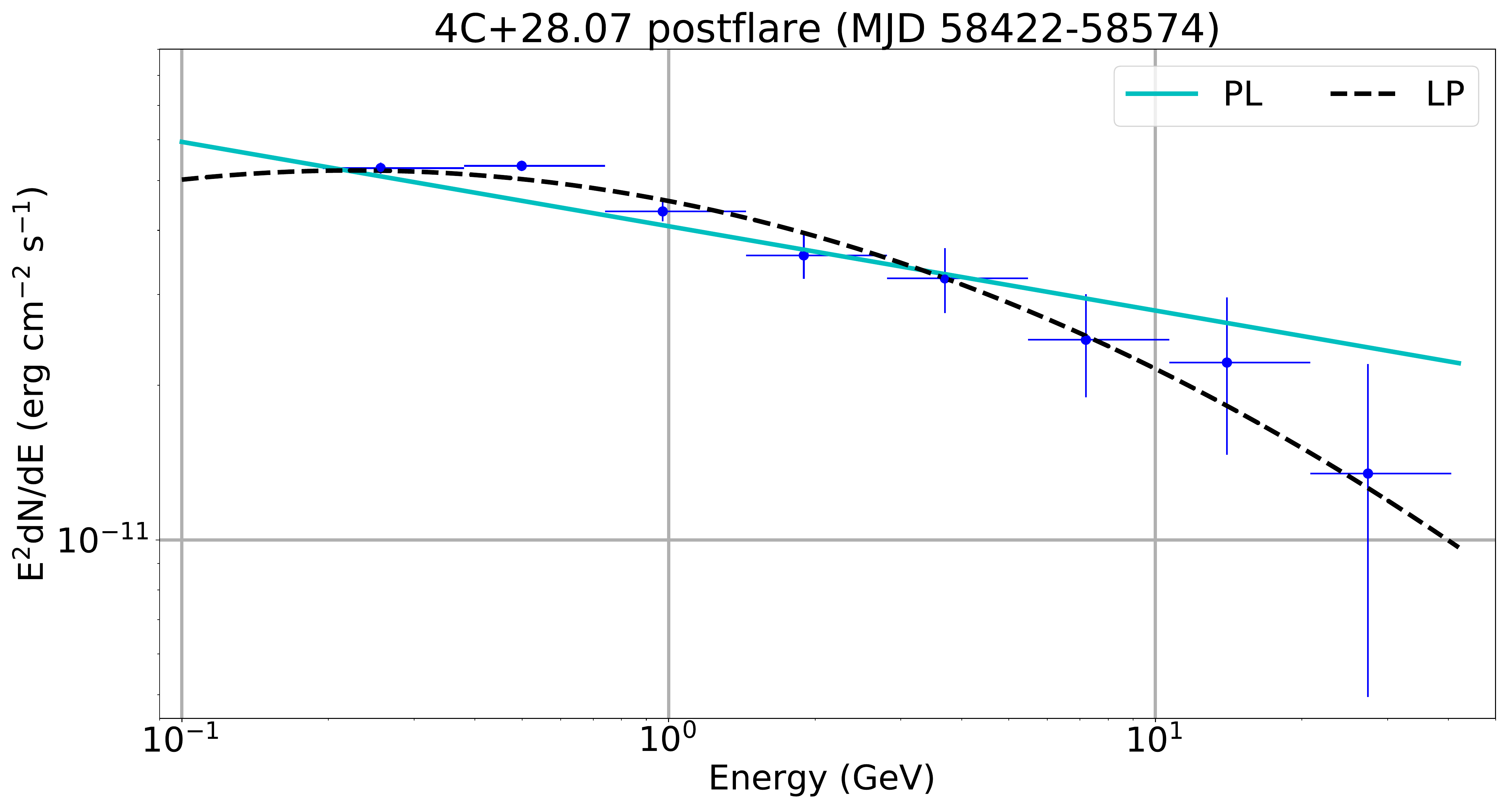}
\caption{$\gamma$-ray SED of different periods of activity (flare, Post-flare) of Flare-C. PL, LP describe the Powerlaw, Logparabola model respectively, which are fitted to data points.}
\label{fig:11}

\end{figure*}

Similar to other flaring states, SED of Flare-C also shows (shown in Figure-\ref{fig:11}) spectral hardening. When the source entered from Flare to Post-flare phase, its spectral index $\Gamma_{PL}$ changed from 2.10$\pm$0.02 to 2.16$\pm$0.02. The modelling parameters have been described in Table-\ref{tab:6}.

The differences in log-likelihood values have also been given in the tables (Table-\ref{tab:4} to Table-\ref{tab:6}) to compare the LP model with power-law model:

\begin{equation}
    \Delta log(\mathcal{L}) = \lvert log(\mathcal{L})_{LP} \rvert - \lvert log(\mathcal{L})_{PL} \rvert
\end{equation}

Where, $log(\mathcal{L})_{LP}$ \& $log(\mathcal{L})_{PL}$ represent the log-likelihood value of log parabola (LP) and power-law (PL) model respectively. 
Table-\ref{tab:7} shows the reduced- $\chi^{2}$ value for two different model. The value of $\Delta log(\mathcal{L})$ and the reduced- $\chi^{2}$ suggest that the LP is better describing the $\gamma$-ray SED than PL.


\section{Multi-wavelength Study}

\subsection{MULTI-WAVELENGTH LIGHT CURVE \& CORRELATION STUDY}
\begin{figure*}[h]
\centering
\includegraphics[height=4.7in,width=8.2in]{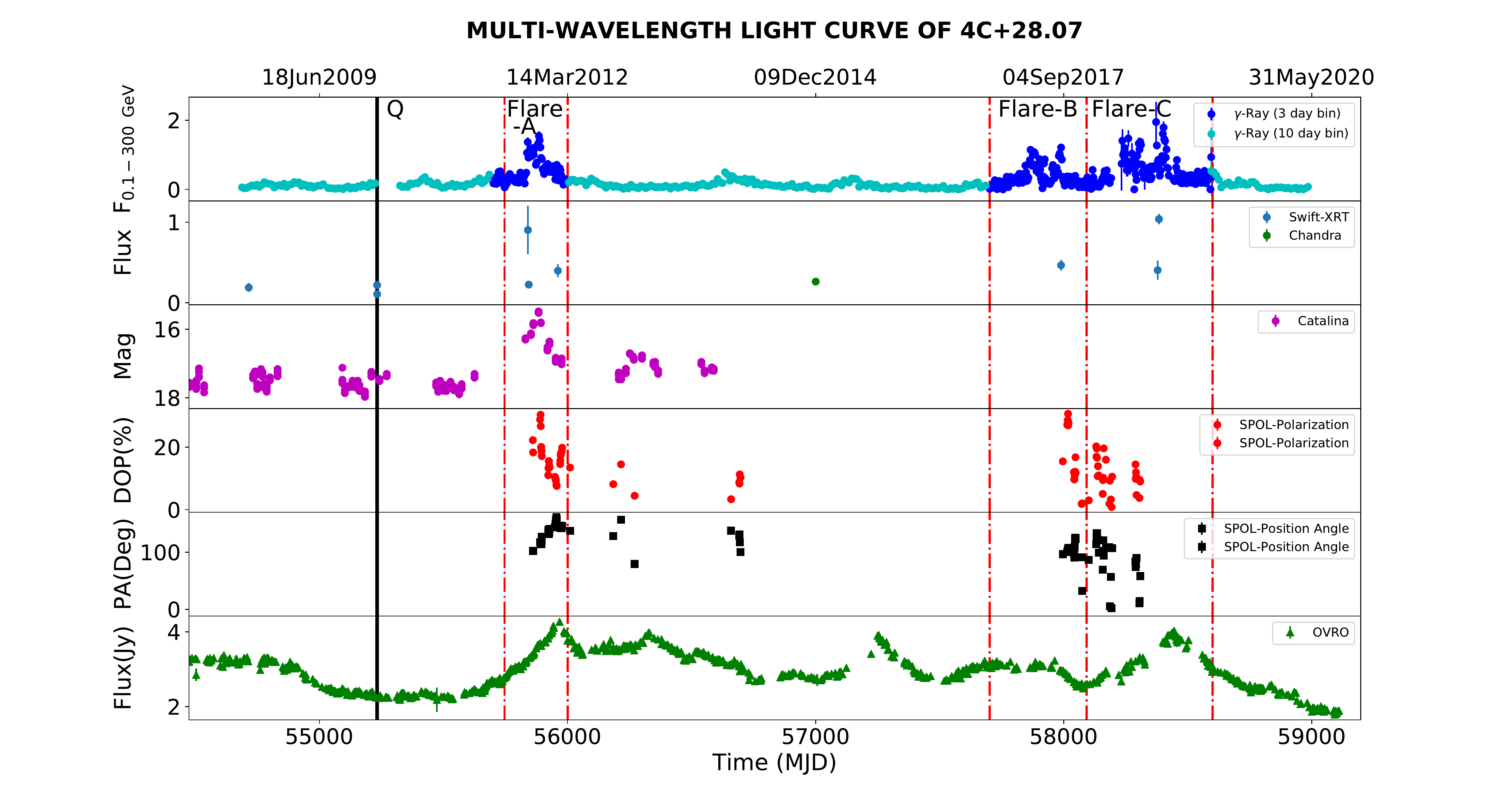}
\caption{Multi-wavelength light curve of 4C+28.07. Top panel shows the Fermi-LAT data with 3 day \& 10 day time bin. Swift-XRT \& Chandra data are shown in second panel in units of $10^{-11}$ erg cm$^{-2}$ s$^{-1}$.}
\label{fig:12}

\end{figure*}

In this section, we have studied the multi-wavelength property (Light curve $\&$ SED) of 4C+28.07 in various bands. Figure-\ref{fig:12} shows the multi-wavelength light curve of this source in $\gamma$-ray, X-ray, optical and radio wavebands. SPOL- polarization \& Position angle data are also shown in the fourth \& fifth panel of the plot, and significant change is observed during all flaring state. A correlation study between these light curves can give an idea about the location of different emission regions in the jet. For this purpose we have done zDCF (Discrete Correlation Function; \citet{1988ApJ...333..646E}) analysis. Unbinned Discrete Correlation Function (UDCF) is given by - 
\begin{equation}
    UDCF_{ij} = \frac{(a_i - <a>)(b_j - <b>)}{\sqrt{ (\sigma_a^2 - e_a^2)(\sigma_b^2 - e_b^2)}}
\end{equation}

where, $a_i$, $b_j$ are the two discrete time series data sets. $\sigma_a$ \& $\sigma_b$ are their standard deviations.
If we bin the above quantity (UDCF$_{ij}$) over M number of pairs for which $\tau - \Delta \tau/2 \leq \Delta t_{ij}  (= t_j - t_i) \leq \tau + \Delta \tau/2$, we will get the Discrete Correlation Function (DCF) -

\begin{equation}
    DCF(\tau) = \frac{1}{M} UDCF_{ij} \pm  \frac{1}{M-1} \sqrt {\sum [UDCF_{ij} - DCF(\tau)]^2}
\end{equation}

\begin{figure*}[h!]
\centering
\includegraphics[height=2.4in,width=3.3in]{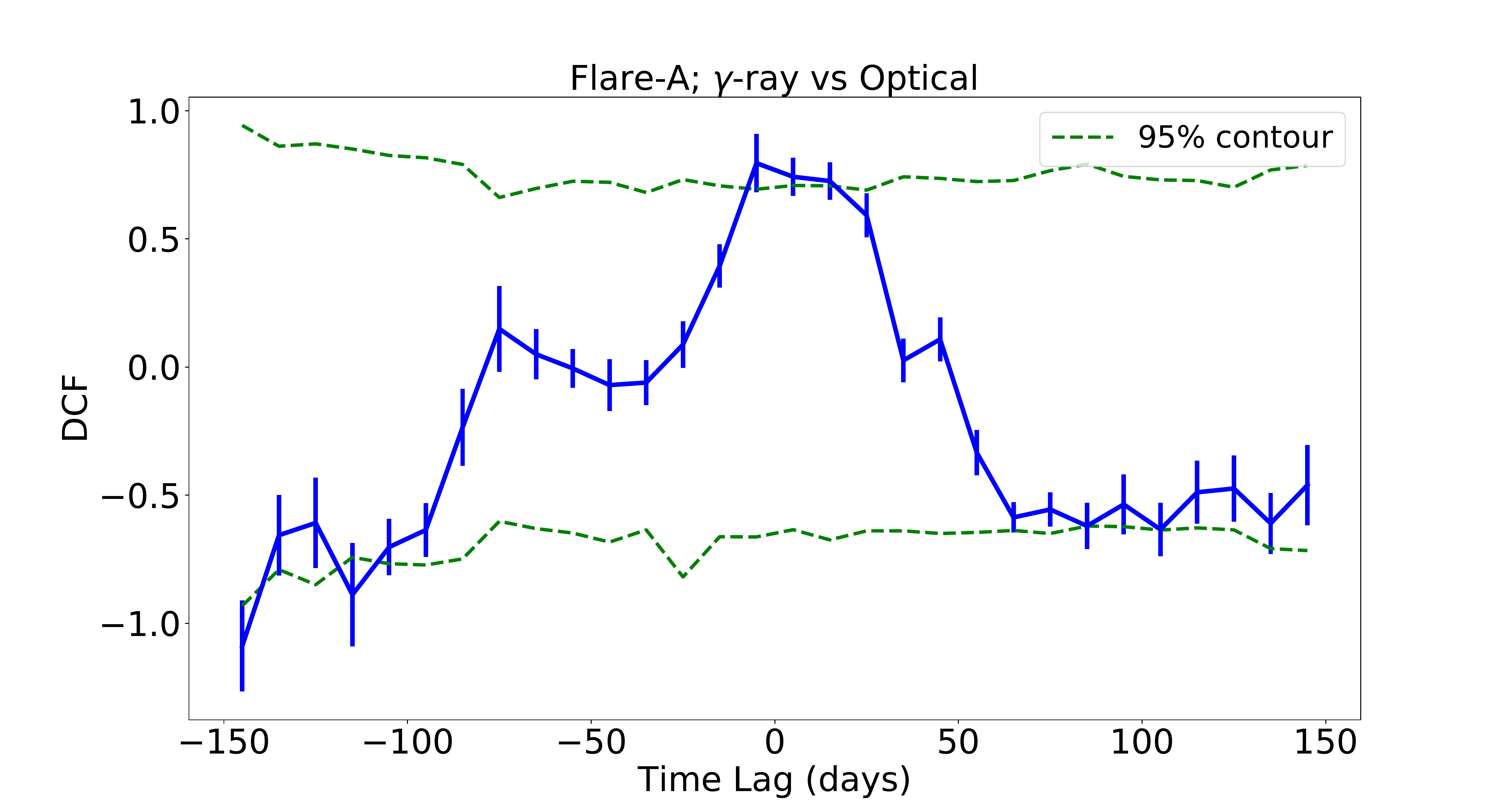}
\caption{DCF plots between $\gamma$ vs Optical data. 95 \% contour is shown in green dashed line.}
\label{fig:13}

\end{figure*}

\begin{figure*}[h!]
\centering

\includegraphics[height=2.4in,width=3.3in]{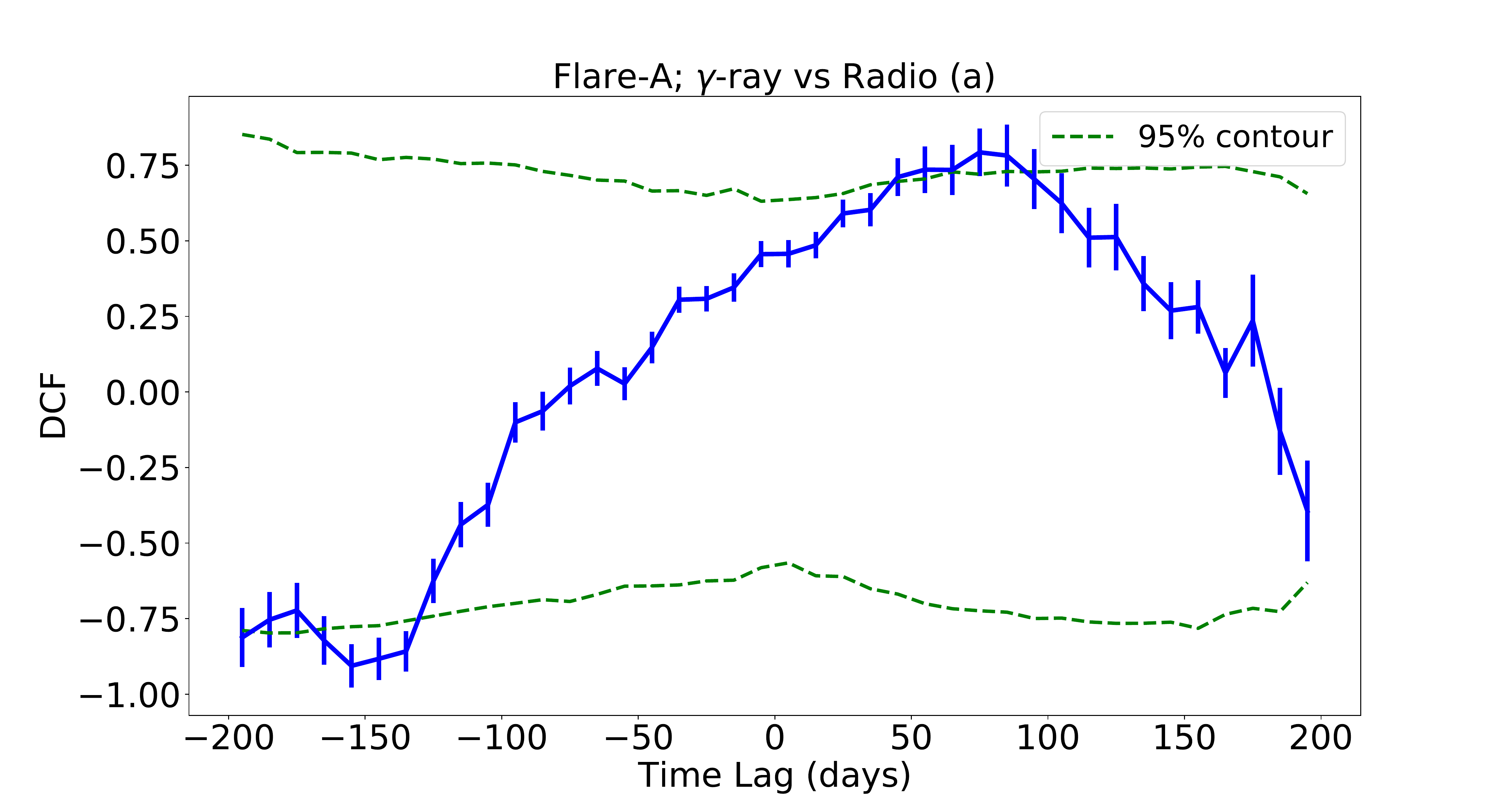}
\includegraphics[height=2.4in,width=3.3in]{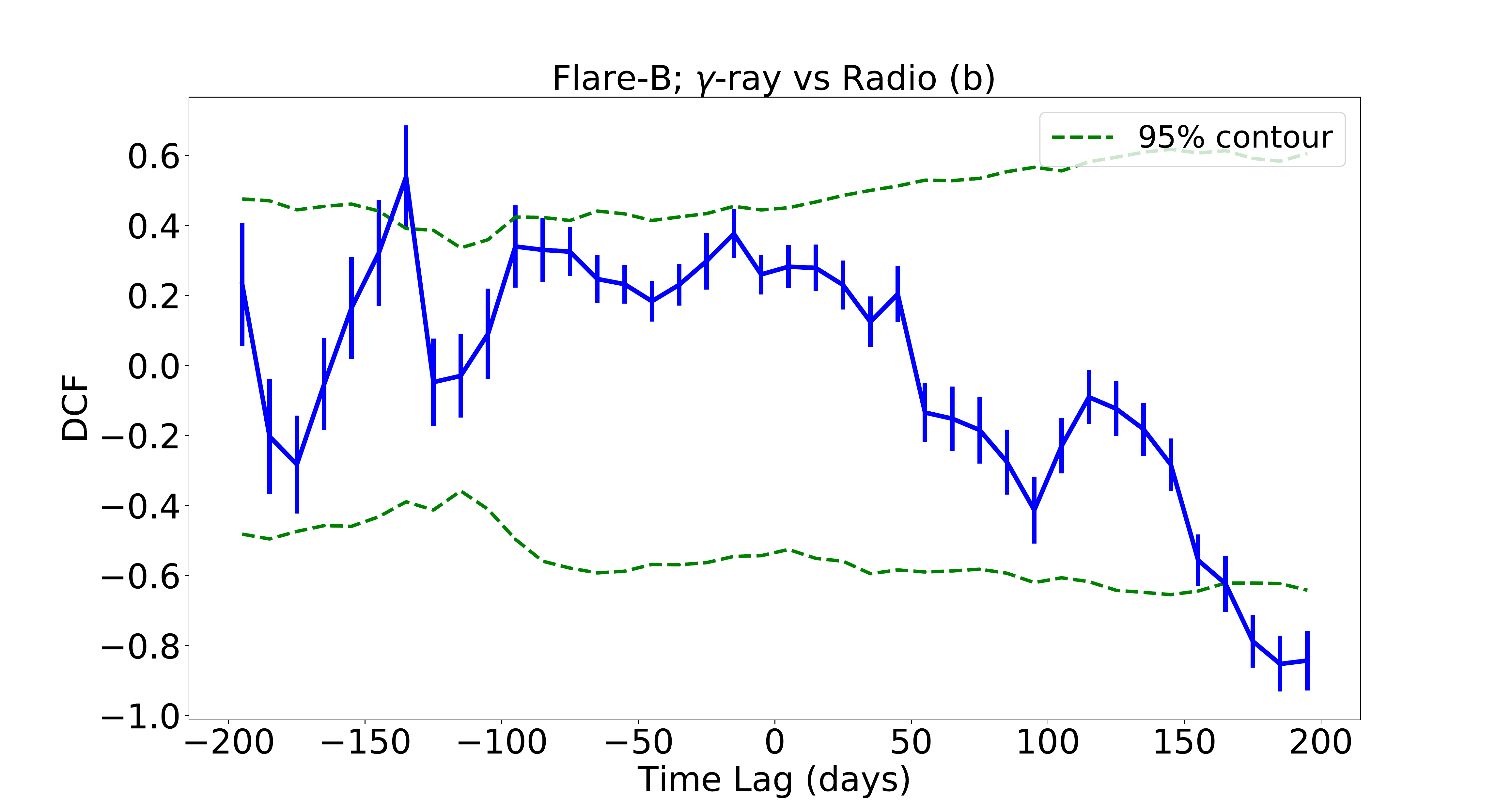}
\includegraphics[height=2.4in,width=3.3in]{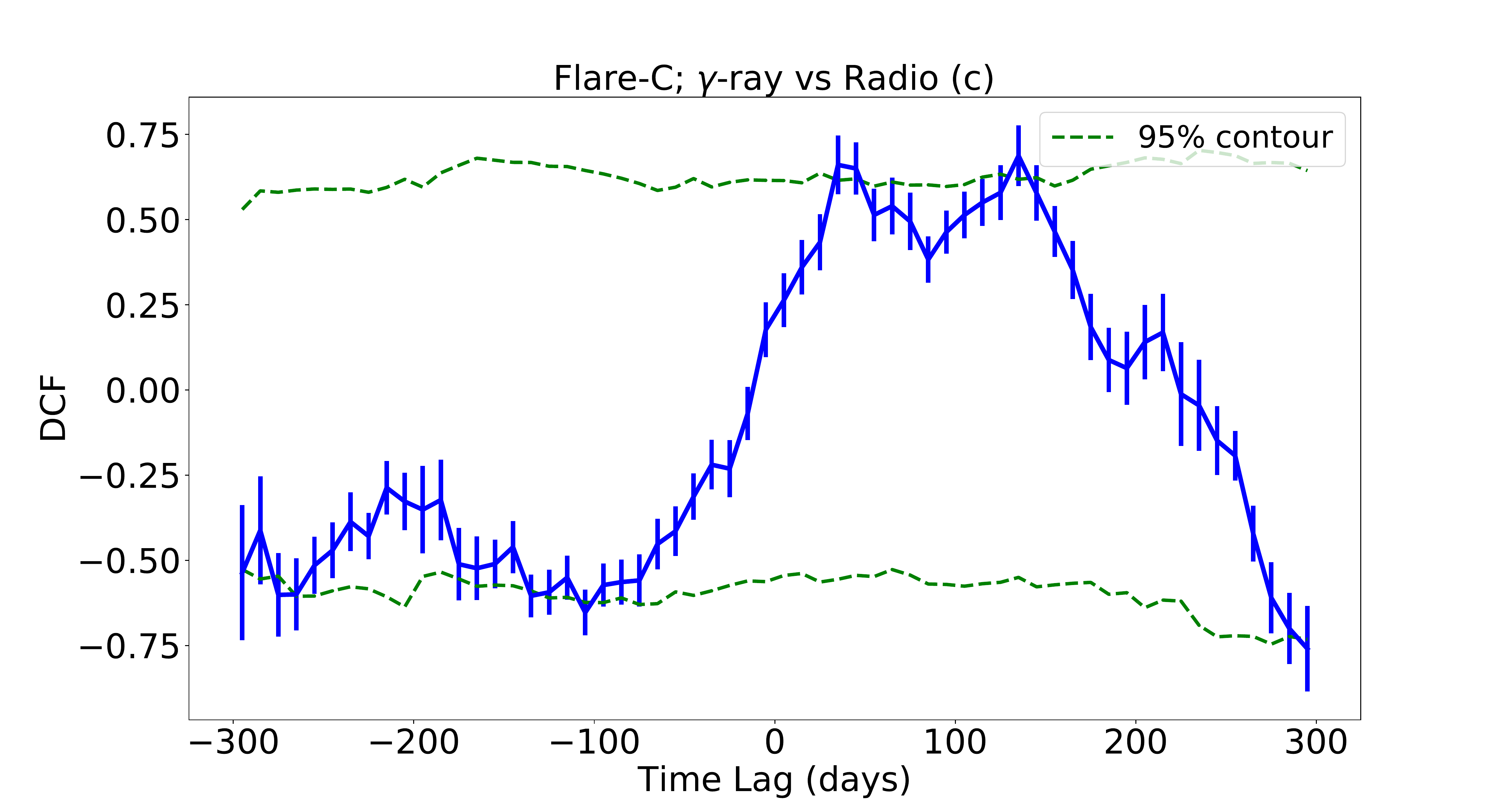}
\caption{DCF plots between $\gamma$ vs Radio data. 95 \% contour is shown in green dashed line.}
\label{fig:14}

\end{figure*}

Figure-\ref{fig:13} and Figure-\ref{fig:14} show the DCF plots between $\gamma$-ray \& optical and $\gamma$-ray \& radio data respectively. Simultaneous $\gamma$-ray \& optical (Catalina) data is available only for Flare-A. From Figure-13, we can see nearly zero time lag ($\sim$ 6 days) between $\gamma$-ray and optical light curves. However, between $\gamma$ \& radio bands, a significant time lag of $\sim$ 70 days is observed (shown in Figure-14a). For Flare-B, there are two distinct peaks in $\gamma$-ray, but no peaks in radio data, as shown in the multi-wavelength light curve in Figure-12. However, in the DCF plot of Flare-B (Figure-14b), only one significant peak ($>$ 95\%) is observed at time lag of $\sim$ -135 days, which is at the edge of the DCF or beyond one third of the total time length of the worst time series \citep{2019ApJ...883..137P}, hence we did not consider this as a proper time lag between $\gamma$-ray \& radio light curve. Thus no significant correlation between $\gamma$-ray \& radio wavelength is found for Flare-B. In $\gamma$-ray region, Flare-C has two distinctive peaks, whereas one clear peak is observed in the radio band. As a result, Figure-14c shows two peaks at time lag $\sim$35 \& $\sim$134 days with DCF value of 0.58 \& 0.63 respectively. 
We did not preform any correlation of X-ray with other wavebands because of lack of sufficient data points in X-ray band. 

To compute the significance of the DCF peaks, we have simulated 1000 $\gamma$-ray light curves using the method mentioned in \citet{2013MNRAS.433..907E}. This method accounts for the general Probability distribution function (Log-normal) of the photon flux. We then computed the DCF between each of the simulated $\gamma$-ray data with optical \& radio data respectively. For each time lag 95\% significant contour is shown in the DCF plots (Figure-\ref{fig:13} - Figure-\ref{fig:14}). 

Here, we have done correlation study for the flaring states only. Earlier, \citet{2014MNRAS.445..428M} performed the correlation analysis for the same source along with several other sources for the first three years (2008 August 4 - 2011 August 12) of Fermi-LAT data and four years (2008 January 1 - 2012 February 26) of 15 GHz OVRO  data set, though no significant correlation was found.

DCF plots suggest that there are significant time lags between $\gamma$-ray \& radio band for Flare-A \& Flare-C. However, no correlation is observed for Flare-B. Flare-A \& Flare-C consist of only one flaring phase, whereas Flare-B consists of two flaring phases (flare-B1 and flare-B2) in its $\gamma$-ray light curve. Significant variability in flux is observed in the $\gamma$-ray band for different flaring phases but no variation is observed in the low frequency (15 GHz radio band) region during this time interval.



	

 In our work, we have used two zone emission model to explain the multi-wavelength SED (for explanation see `SUMMARY \& DISCUSSION' section). The inner blob (Blob-I) emits radiation via synchrotron, SSC (Synchrotron Self Compton) \& EC (External Compton) processes and can explain the X-ray and $\gamma$-ray part of the spectrum, whereas the outer blob (Blob-II) contributes to the SED in radio \&  partially in X-ray frequency via synchrotron \& SSC process respectively, as there are no external photons (from disk/BLR/DT) available for EC emission. The size of the BLR \& DT region can be calculated by the simple scaling relations : $R_{BLR} \sim L_{Disk,45}^{0.5} \times 10^{17}$ cm and $R_{DT} \sim 2.5 L_{Disk,45}^{0.5} \times 10^{18}$ cm \citep{2009MNRAS.397..985G}. We have  approximated these sizes as 9.8$\times 10^{17}$ cm \& 1.1$\times 10^{19}$ cm respectively.  Blob-II is located beyond these regions (BLR \& DT), hence external seed photons from BLR and DT are not abundant at the location of Blob-II for EC emission.

\subsection{MULTI-WAVELENGTH MODELLING}

We have modeled the multi-wavelength SED of 4C+28.07 with the publicly available code: `GAMERA' \footnote{http://joachimhahn.github.io/GAMERA} \citep{Hahn:2016CO} which considers the time dependent spectral evolution while modelling the SED. This code computes the synchrotron and IC component of radiation by the given electron spectrum, $N(E,t)$. This $N(E,t)$ is calculated by solving the following continuity equation - 

\begin{equation}\label{eq:7}
 \frac{\partial N(E,t)}{\partial t} = Q(E,t)-\frac{\partial }{\partial E}(b(E,t)N(E,t))-\frac{N(E,t)}{\tau_{esc}(E,t)}   
\end{equation}

Here, $Q(E,t)$ is the injected electron spectrum. $b(E,t)$ \& $\tau_{esc}(E,t)$ are the energy loss rate \& escape time of the electrons respectively. `GAMERA' uses the full Klein Nishina cross section to compute the IC radiation \citep{1970RvMP...42..237B}. Log-parabolic functional form of $Q(E,t)$ is assumed to solve the transport equation as most of the $\gamma$-ray SED flaring states are best described by this model (\citet{2004A&A...413..489M}; \citet{2020ApJS..248....8D}).

The value of Lorentz factor $\Gamma$ is assumed to be 12.7 following \citet{2010A&A...512A..24S}. Doppler factor ($\delta$) of the jet is chosen as 25, which is a typical value for FSRQ type blazars (\citet{2017MNRAS.466.4625L}; \citet{2009A&A...507L..33P}; \citet{2018ApJ...852...45W}).

BLR, CMB \& accretion disk photons are considered as seed photons to compute the EC emission component for Blob-I. SSC component for both the blobs have been also estimated to model the source.

Energy density of BLR photons in the comoving frame of the jet is given by - 

\begin{equation}\label{eq:8}
 U^\prime_{BLR} = \frac{\Gamma^2 f_{BLR}L_{Disk}}{4\pi cR^2_{BLR}}   
\end{equation}


where, $f_{BLR}$ is the ratio of BLR photon energy density to accretion disk photon energy density, assumed to be $\sim$ 0.1. $R_{BLR}$ is the radius of the BLR region, which is calculated by using the scaling relation \citep{2009MNRAS.397..985G} - $R_{BLR} = 10^{17}L^{1/2}_{d,45}$. $L_{d,45}$ is the disk luminosity in unit of $10^{45} erg/sec$. From \citet{1997MNRAS.286..415C}, \citet{2003ApJ...593..667M}; \citet{2014ApJ...788..104Z} we can assume that accretion disk luminosity is approximately 10 times higher than the BLR luminosity ($L_{Disk} \sim 10 \times L_{BLR}$). In our study, we have taken the value of $L_{BLR} = 10^{45.39}$ erg/sec from \citet{2014MNRAS.441.3375X}. We have used the BLR temperature as $1.0 \times 10^{4}$ K \citep{Dermer:1225453}.

The following relation is used to calculate the energy density of the disk photons at the location of Blob-I \citep{Dermer:1225453} - 

\begin{equation}\label{eq:9}
 U^\prime_{Disk} = \frac{0.207 R_{g}L_{Disk}}{\pi c Z^3 \Gamma^2}   
\end{equation}

where, $R_g$ $(= \frac{G M_{BH}}{c^2})$ is the gravitational radius. The mass of the central engine ($M_{BH} \sim 10^{9.22} M_\odot$) is taken from \citet{2014MNRAS.441.3375X}. `Z' is the distance between $\gamma$-ray emission region (Blob-I) and central engine of the system, which is estimated to be 9.47$\times10^{17}$ cm using this equation $Z = \frac{2 \Gamma^{2} c t_{var}}{1+z}$, $t_{var}\sim$ 2.5 days. Equation-\ref{eq:9} gives the value of $U^\prime_{Disk}$ as 9.0$\times10^{-8}$ erg cm$^{-3}$. The values used in this work are $U _{Disk}^{\prime}$ = $7.0 \times 10^{-8}$ erg cm$^{-3}$ \& $T_{Disk}^{\prime}$ = $6.0 \times 10^4$ K, which are comparable to the values found for other FSRQs. The sharp fall in optical-UV region is explained by the direct disk emission. 
The energy density of disk photons is much lower than the energy density of BLR photons, hence EC emission by disk photons is not important. 


Apart from these, we also constrain the radius of Blob-I ($R_I$) by using the following relation - 

\begin{equation}\label{eq:10}
 R_I \leqslant \frac {c t_{var} \delta}{1+z}.   
\end{equation}

The value of variability time ($t_{var}$) is used as $\sim$ 2.5 days, which gives the upper limit of the emission region : $R_{I} \sim 4.7 \times 10^{16}$ cm. The above formula is just an approximation \& there are several effects that may introduce error in determining $R_I$ \citep{2002PASA...19..486P}. We have used the radius of Blob-I  as $2.0 \times 10^{16}$ cm, which gives best fit to the SED data. We have computed the distance between Blob-I \& Blob-II by the following equation \citep{2014MNRAS.441.1899F} - 
\begin{equation}\label{eq:11}
 \Delta r_{I-II} = \frac {\beta_{app} c \Delta t_{obs}}{(1+z) \sin{\theta}}.   
\end{equation}

Where, $\beta_{app} = 18.8$ is the apparent speed of jet. $\Delta t_{obs}$ \& $\theta$ are the observed time lag \& viewing angle of the source respectively. Viewing angle is chosen as 3.4$^\circ$ \citep{2009A&A...494..527H}. By using the above values we have computed $\Delta r_{I-II}$ as 2.79$\times 10^{19}$ cm and 1.30$\times 10^{19}$ cm for flare phases of Flare-A and Flare-C. After that, we have calculated the radius of Blob-II under conical jet approximation, which gives the value of 1.44$\times 10^{18}$ cm \& 6.94$\times 10^{17}$ cm for the above flare phases (Flare-A \& Flare-C respectively). However, these high values of radii increase the required electron jet power in the emission region (Blob-II) and as a result the total jet power exceeds the Eddington luminosity ($L_{edd} = 2.29 \times 10^{47}$ erg/sec) of the source. Hence, in our model we used optimum value of the Blob-II radius as 6.5$\times 10 ^{17}$ cm for all the flares \& quiescent state.   





\begin{figure*}[h!]
\centering

\includegraphics[height=2.5in,width=3.2in]{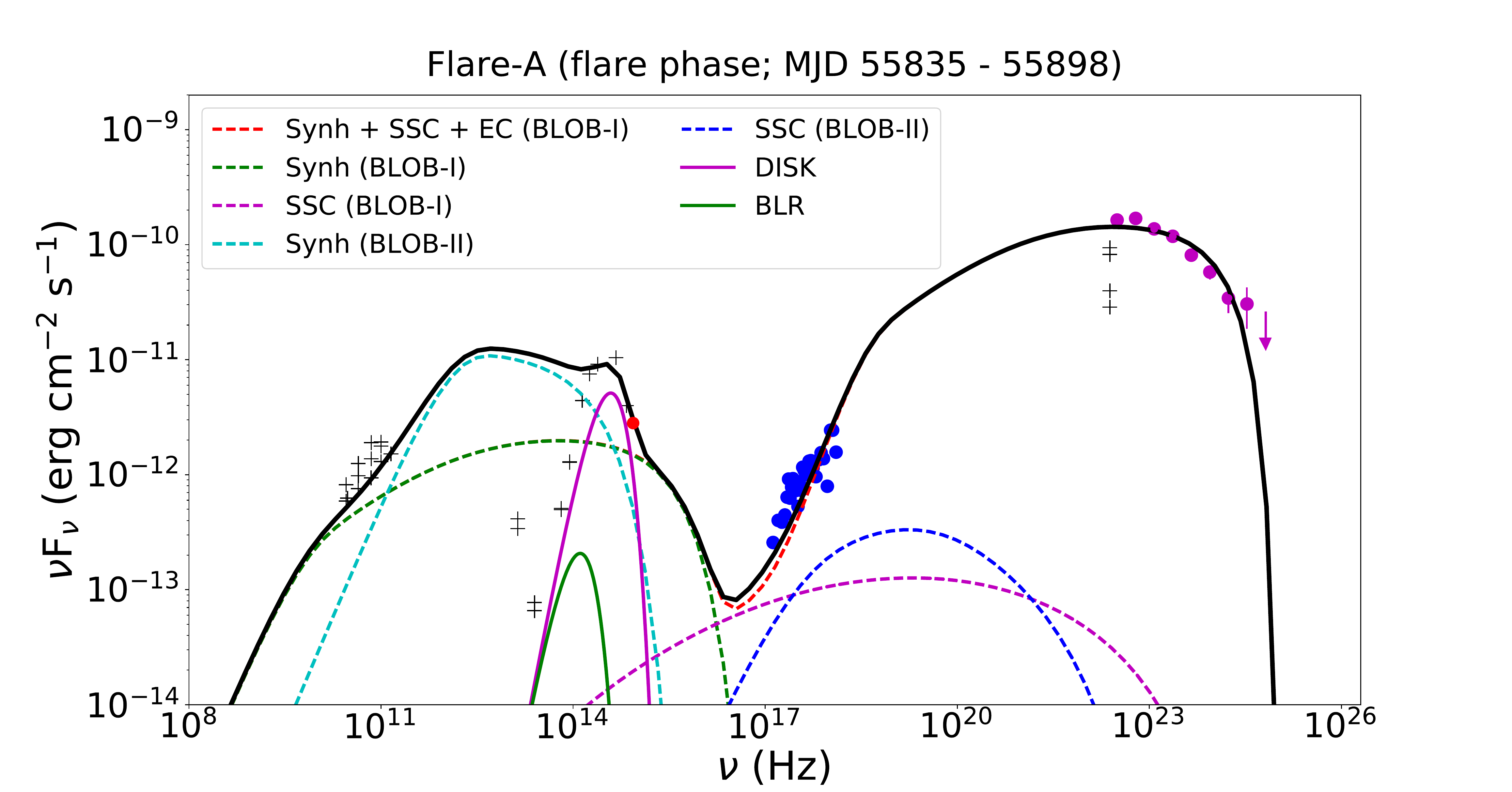}
\includegraphics[height=2.5in,width=3.2in]{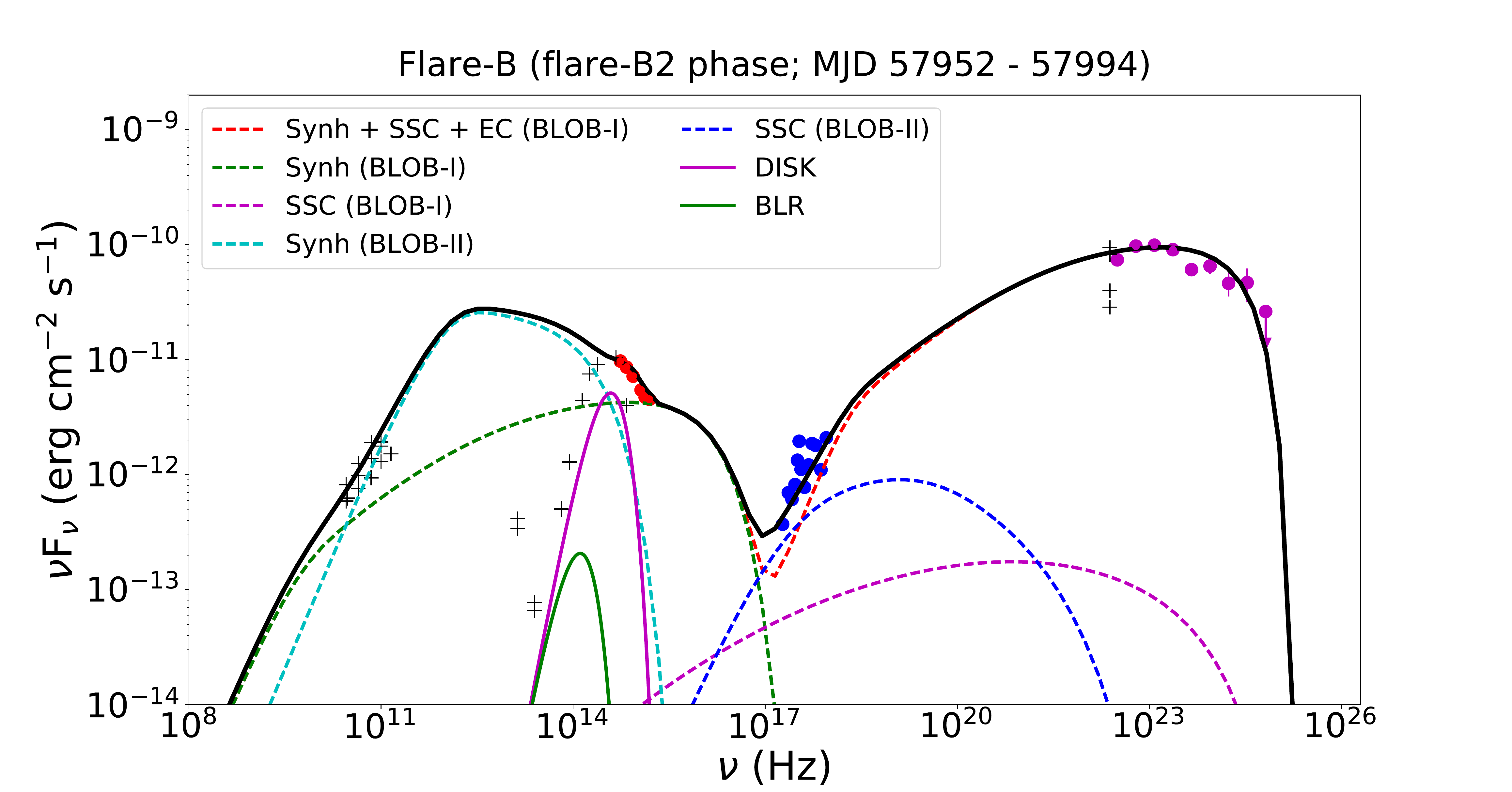}
\includegraphics[height=2.5in,width=3.2in]{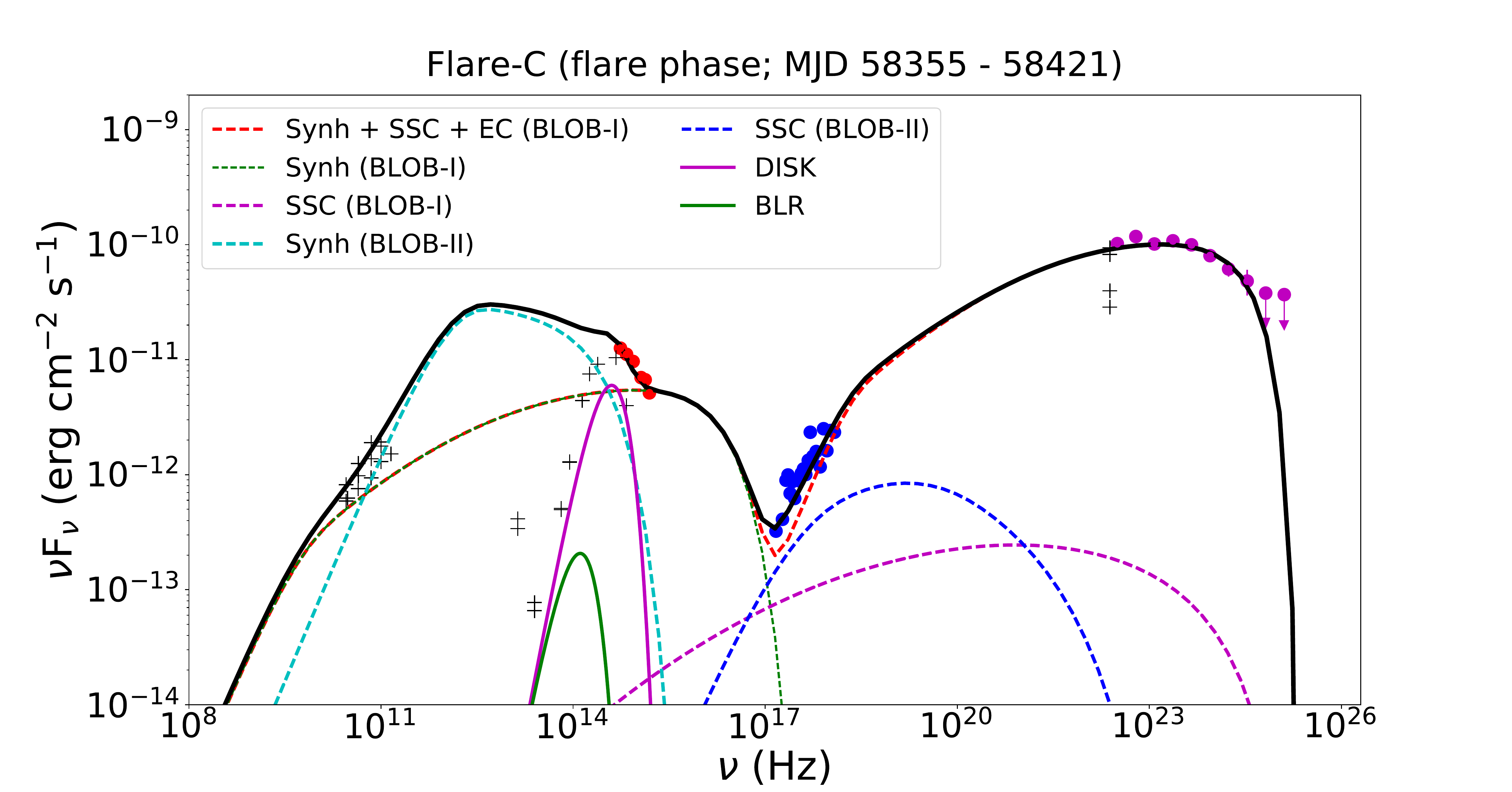}
\includegraphics[height=2.5in,width=3.2in]{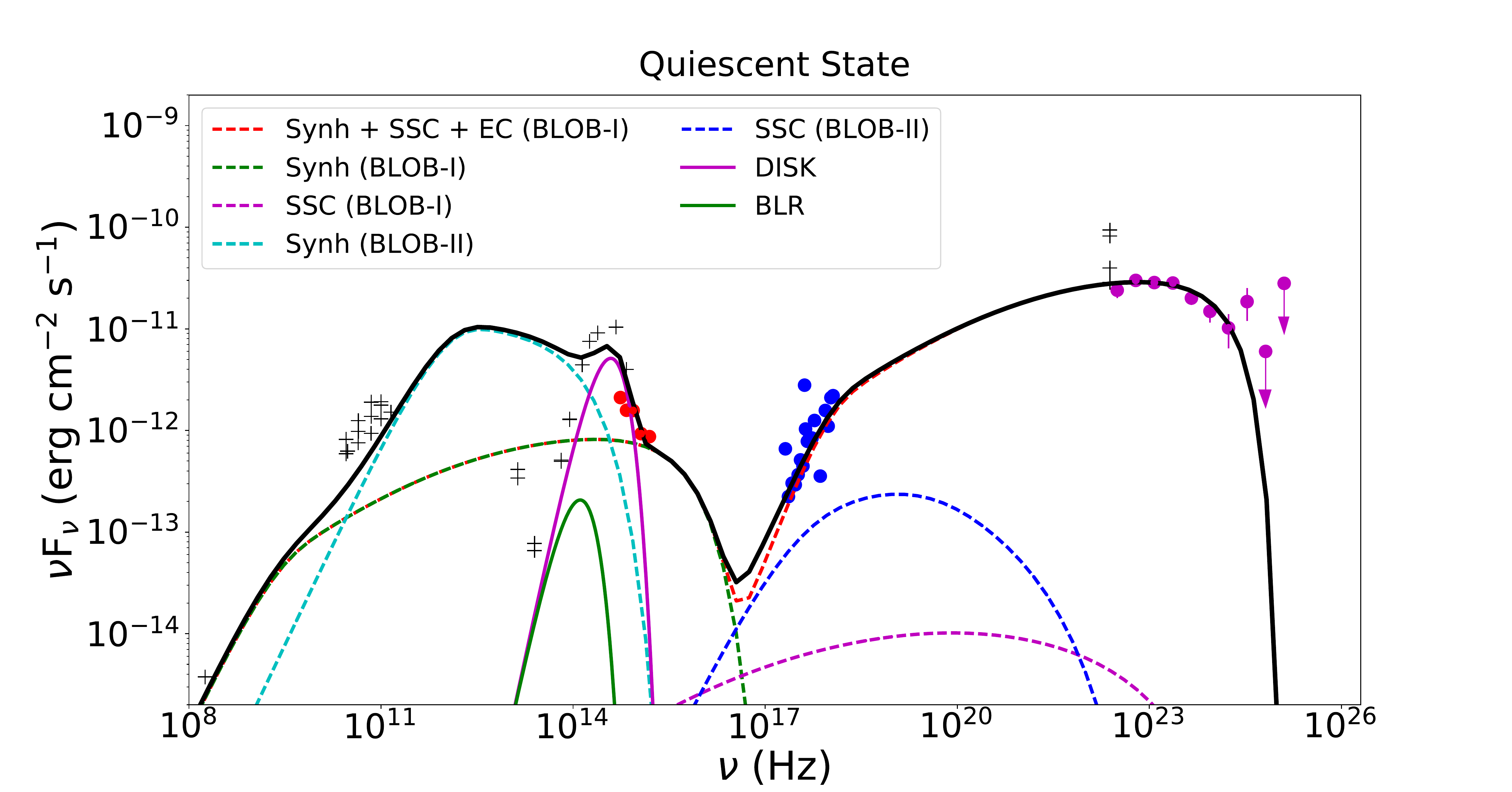}
\caption{Two zone model fits of the multi-wavelength SEDs. Emission processes of Blob-I \& Blob-II have been shown in different colors. Disk \& BLR emission are also illustrated in solid magenta \& green color respectively.}
\label{fig:15}

\end{figure*}

\begin{figure}
\centering
\includegraphics[scale=0.4]{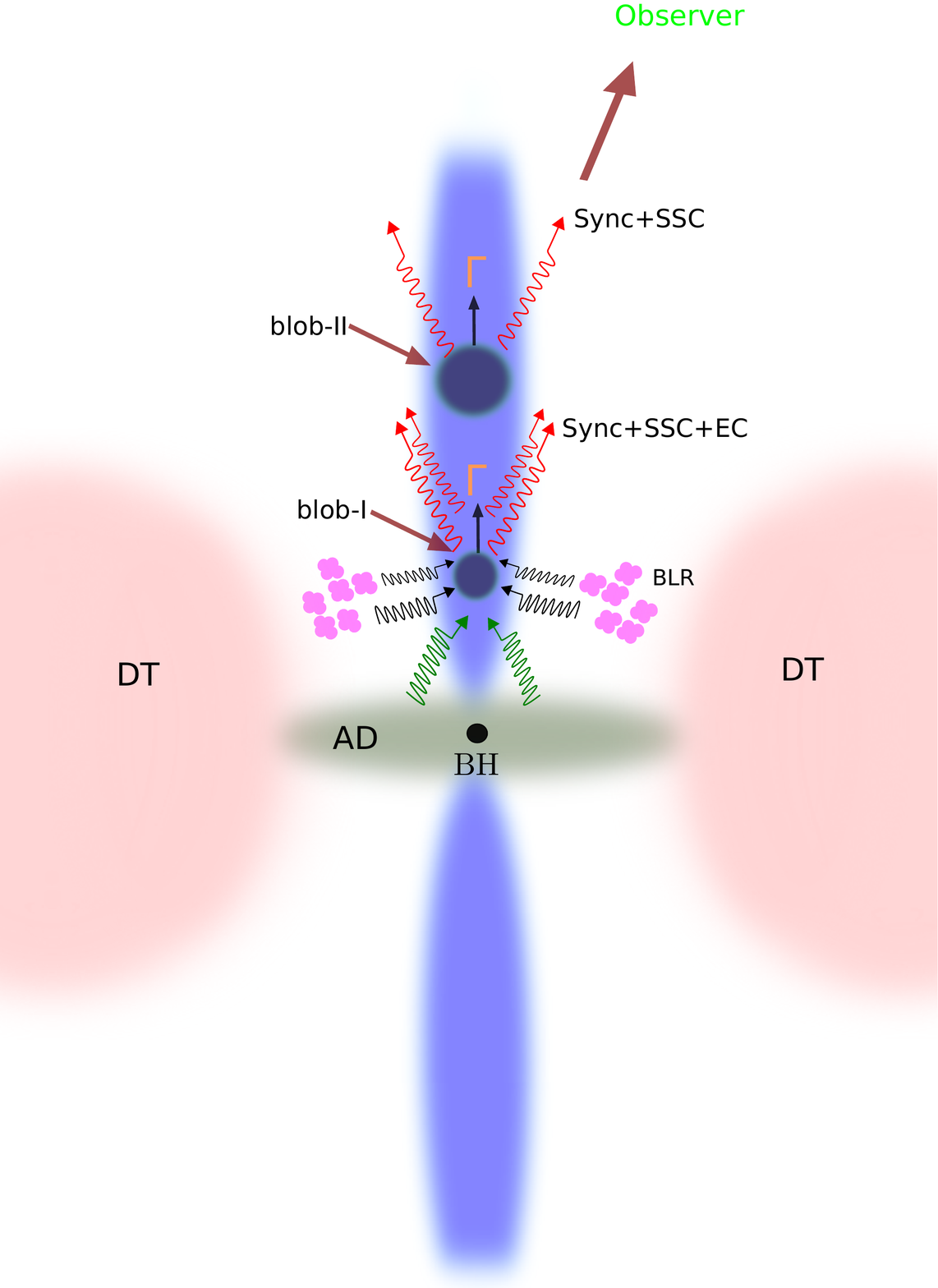} 
\caption{Schematic representation of two blob model in the jet used for the broadband SED modeling. Image is not in scale. AD: Accretion Disk; BH: Black Hole; BLR: Broad Line Region; DT: Dusty Torus. 
    }
\label{fig:cartoon}
\end{figure}

After constraining the above parameters, we have modelled all the multi-wavelength SEDs of flaring phases except Flare-B1 (due to non-availability of simultaneous X-ray \& optical-UV data) and one quiescent state, which are shown in Figure-\ref{fig:15}.  We have also shown the non-simultaneous archival data with cyan plus point. A pictorial representation of our model is shown with a cartoon diagram in Figure-\ref{fig:cartoon}. This represents the case of Flare-A where Blob-II is around 9 pc away from the SMBH, beyond the Dusty Torus (DT) region. For Flare-C though the situation is very similar, Blob-II is near the boundary of the DT.

There are several free parameters in our model, e.g: spectral index ($\alpha$), curvature index ($\beta$) \& normalization factor ($l_0$) of the injected log parabolic electron spectrum, minimum ($e_{min}$) \& maximum ($e_{max}$) energy of the injected electrons, magnetic field ($B$) of the emission regions. 

Total required jet power of both blobs are computed by the following equation - 

\begin{equation}\label{eq:12}
 P_{tot} = \pi R^2 \Gamma^2 c \sum_i (U^\prime_{e,i}+U^\prime_{B,i}+U^\prime_{p,i})   
\end{equation}

where, i runs over Blob-I and Blob-II. Energy density of electrons, magnetic field \& cold protons in the co-moving frame are represented by $U^\prime_e$, $U^\prime_B$ \& $U^\prime_p$ respectively. Energy density of electrons and magnetic field are given by - 
\begin{equation}\label{eq:13}
 U^\prime_{e,i} = \frac{3}{4 \pi R_{i}^3}\int_{e_{min}}^{e_{max}} E Q_{i}(E) dE  
\end{equation}
and,

\begin{equation}\label{eq:14}
 U^\prime_{B,i} =  \frac{B^2_{i}}{8 \pi}.
\end{equation}

$Q(E)$ is the injected electron (log-parabolic form for both blobs) spectrum in the jet - 

\begin{equation}\label{eq:15}
   Q(E) = l_0\Big(\frac{E}{E_{ref}}\Big)^{-(\alpha+\beta\log(E/E_{ref}))}
\end{equation}
$l_0$, $\alpha$, $\beta$ are the normalization factor, spectral \& curvature index respectively. $E_{ref}$ is the reference energy, which is fixed at 90.0 MeV. We have assumed, 10:1 ratio of the electron-positron pair and cold protons to compute the value of $U^\prime_p$.

In our study, we have calculated the total required jet power of all the flaring states and one quiescent state and found the maximum value as 9.64$\times$10$^{46}$ erg sec$^{-1}$ for Flare-A. The detailed description of multi-wavelength SED modelling parameters have been elucidated in Table-\ref{tab:8}.

\section{Summary and Discussion}
The blazar 4C +28.07/BZQJ0237+2848 has not been studied before with multi-wavelength SED modelling. Here, we present the first comprehensive temporal and spectral study on this source.
\par
We have analyzed the Fermi-LAT light curve of FSRQ 4C+28.07 in 10 days binning with time span of $\sim 12$ years (MJD 54682 - 59000) and found three distinct Flares: Flare-A, Flare-B \& Flare-C. These Flares have been further analyzed in 3 days time bin, which are illustrated in Figure-1. Due to irregular \& large error-bar in photon counts, we have restricted our analysis in 3 day binning only. 
In three day binned light curve, we have applied Bayesian block method to identify the various phases within each flare. The following phases have been identified pre-flare, flare and post-flare. The flux values above 5$\sigma$ considered to be part of the high active phases e.g. flare-A, flare-B1, etc. We have also collected the broadband data from various ground and space based telescopes such as Swift, Steward, Catalina and OVRO (15 GHz). In Swift-XRT, we do not have much observations and hence we could not perform any temporal study. However, the spectrum from a few observations available during flare phases of Flare-A, Flare-B, and Flare-C is used for the broadband SED modeling. 

\par
This source has been included in F-GAMMA program and observed in the frequency range of 2.64 GHz to 43 GHz between 2007 and 2015 \citep{2019A&A...626A..60A}. This source has also been  monitored by the MOJAVAE survey \citep{2019ApJ...874...43L} since 1994. We have used the OVRO data from \citep{2011ApJS..194...29R} in 15 GHz radio band, which covers the period of Fermi-LAT observations.
\par
We have performed the correlation study among various optical and radio band emission at 15 GHz with $\gamma$-ray. 
They show a very small time lag (order of a day) between $\gamma$-ray and optical emission with very high correlation coefficient ($\sim$80$\%$). However, a significant time lag of the order of $\sim$70 days and 35 days is observed between $\gamma$-ray and radio emission at 15 GHz  for Flare-A and Flare-C with more than 60$\%$ significance. The Flare-B does not show any correlation between $\gamma$-ray with radio band emission. To model the broadband SED from radio to $\gamma$-ray for Flare-A and Flare-C, we have used two zone emission model responsible for the radio and $\gamma$-ray emission separately. A cartoon diagram for the two zone model is shown in Fig. 16, though the image does not represent the correct relative scale. In Flare-A, these emission regions are separated by $\sim$9 pc and in Flare-C they are separated by $\sim$4 pc along the jet axis. Similar results were also observed in other FSRQ Ton 599 by \citet{Prince2019} where a separation of $\sim$5 pc was found between the $\gamma$-ray and radio band emission at 15 GHz. 
One way to explain the separation between $\gamma$-ray and radio emission could be the shock in jet model (\citet{Marscher1985}, \citet{Valtaoja1992}).
The delay in emission at lower frequency can be caused by high opacity at the base of the jet under the shock in jet model, and the lower frequency emission can be absorbed by the synchrotron self-absorption. In shock in jet model, a shock is formed at the base of the jet where the jet is optically thick at the radio frequency but transparent for the high energy. So, the broadband emission produced at the base of the jet will be observed first at optical/$\gamma$-ray and later at radio when the jet becomes optically thin for the radio emission. The connection between $\gamma$-ray, optical variability and parsec scale radio emission of 3C 345 was studied earlier (\citealt{Schinzel2012Jan}). The authors assumed that the emissions in radio, optical and $\gamma$-ray frequencies are simultaneous. They suggested that the $\gamma$-ray emission could be produced by SSC mechanism in a region of the jet that extends up to $\sim$23 pc from the VLBI core, which is beyond the BLR and much more than 1 pc away from the central engine. We tried to fit our SED for 4C+28.07 with a single zone SSC model, where the emission zone is located outside the BLR, several parsecs away from the black hole. We found that it requires super-Eddington luminosity to fit the multi-wavelength data. Moreover, if we keep the blob size $2\times 10^{16}$ cm to explain the observed $\gamma$-ray variability, then we cannot fit the radio data.
\par

A detailed discussion of radio follow-up of $\gamma$-ray emission from PKS 1510-089 at different radio frequencies is provided in \citet{Orienti2013Jan}. Two major $\gamma$-ray flares were detected from PKS 1510-089 in 2011 July and October. After 2011 September a huge radio burst was also detected, first in millimetre, after some time delay in centimetre and also in decimetre wavelengths. This observation of radio emission suggests formation of shock and its evolution, due to expansion of emission region and radiative cooling of electrons. 
\par
The study done in \citet{Aleksic2014Sep} discusses about EC with infrared photons from dusty torus and also the spine-sheath model where the $\gamma$-ray emission region is placed at the radio core to explain the broadband SED. The authors suggested a common origin of millimeter radio and high energy $\gamma$-ray emission.
In a more recent paper by \citet{2021A&A...648A..23H}, 
 the authors discusses that the emission region of very high energy $\gamma$-rays from PKS 1510-089 is located 50 pc away from the black hole down the jet. During the $\gamma$-ray flare VLBI observations at 43 GHz revealed a fast moving knot interacting with the standing jet feature. In their work the very high energy $\gamma$-ray spectrum showed attenuation consistent with EBL absorption, which supports the speculation that the $\gamma$-rays are produced outside the BLR.  
 \par

In our study of blazar 4C+28.07, there is neither any detection of very high energy $\gamma$-rays by MAGIC or HESS nor any detection of simultaneous radio knot emission, therefore we cannot argue that the $\gamma$-ray and radio emissions originate from the same region several parsecs away from the black hole. Moreover, the BLR photons are not available there for efficient production of $\gamma$-rays by EC mechanism. If there is a source of seed photons several parsecs away from the black hole or the radio emission were near the edge of the BLR then one could argue in favour of a single zone emission model. 



Above arguments suggest that, multi-zone emission modeling is required to explain the broadband spectrum from radio to high energy $\gamma$-ray. Therefore, we consider the two emitting zones located at two different location along the jet axis and separated by a distance of $\sim$ 9 pc for Flare-A and $\sim$ 4 pc for Flare-C.
 
 For Flare-B, there is no variability observed in radio band as compared to $\gamma$-ray band. In this case, plasma instabilities could have prevented production of significant radio emission that could relate to $\gamma$-ray activity up-stream or another scenario was shown in \citealt{Schinzel2012Jan}, where $\gamma$-ray emission is shown to be produced at parsec scales along the jet, as well as the presence of fast $\gamma$-ray variability, but this interpretation would further complicate a more simplistic two-zone model.
 
\par
We have discussed in section 5.2 that the radius of the outer blob used in our work is smaller than the radius obtained from conical jet approximation.
This is done under the assumption that the outer blob does  not cover the entire cross-section of the conical jet, otherwise the required jet power is very high. In \citet{2021JHEAp..29...31P} two zone modeling has been done to explain orphan $\gamma$-ray flares from three FSRQs. 
There the first emission region is assumed to be the production site of $\gamma$-rays. This region is located near the dusty torus region for PKS 1510-089 and 3C 279 and its radius is assumed to be smaller than the cross-section of the jet under conical jet approximation to explain the variability observed in the $\gamma$-ray data during orphan $\gamma$-ray flares.
\par
The multi-zone emission modeling is done before also for many other blazars, where the high energy $\gamma$-ray emission could not be explained by single zone emission model. A study by \citet{Cerruti2017} on blazar PKS 1424+240 suggests that to explain the high energy $\gamma$-ray the one zone emission model is not sufficient and a very high Doppler factor above 250 is required in this case, which is unrealistic. Due to this reason they have modelled the broadband SED with the two zone emission model, which allows a reasonable value of Doppler factor, $\delta$ = 30. Recent study of simultaneous $\gamma$-ray flare and neutrino event from TXS 0506+056 by  \citet{Xue2019} also indicates that two zone emission model is required to explain the broadband SED and the emission of neutrino. In their model one emitting zone is located close to the BLR, where the $\gamma$-rays and neutrinos are produced in p$\gamma$ interaction of high energy protons with the BLR photons, and the second emission zone is responsible for the optical and X-ray emission through  synchrotron and SSC (synchrotron self Compton). In \citet{Prince-2019}, they have used two emitting zones to explain the broadband SED of FSRQ PKS 1510-089, where the inner zone is responsible for the optical and $\gamma$-ray emission through synchrotron and EC scattering of BLR photons, and the outer zone is the production site of X-ray emission by EC scattering of dusty torus photons.

Most of the Flares studied in the present work have different activity phases such as Pre-flare, flare, Post-flare etc. We have fitted each flare phase with sum of two exponential functions to compute the rising ($T_r$) and decay ($T_d$) time for different peaks (P1, P2 etc.). Spectral analysis (SEDs) in $\gamma$-rays have also been done for different phases of activity, which are fitted with two different models: Power-law (PL) and Log-parabola (LP). In our study, for every cases (e.g. Flare \& Quiescent states) LP is the best fitted model. 

We define an asymmetry parameter $\zeta$ for different peaks, given by \citep{2010ApJ...722..520A} - 

\begin{equation}\label{eq:16}
 \zeta = \frac{T_d-T_r}{T_d+T_r}   
\end{equation} 

In our work, we have found that 2 peaks have $T_r > T_d$ ($\zeta < -0.3$), 1 peak has $T_r < T_d$ ($\zeta > -0.3$) and 7 peaks have $T_r \sim T_d$ ($\lvert \zeta \rvert \leqslant 0.3$). The last case 
 represents symmetric temporal evolution, which can be explained by 
 perturbation in the jet or a dense plasma blob passing through a standing shock front in the jet region \citep{1979ApJ...232...34B}.
\par
In Table \ref{tab:8} the parameter values for the two zone modelling
 are given. The total jet power required is always lower than the Eddington luminosity of this source $2.29\times 10^{47}$ erg/sec.
 
 \section{ Conclusions}
 
 The FSRQ 4C +28.07 has not been studied before over a long time period with multi-wavelength data. 
 Due to the availability of long term radio and $\gamma$-ray data from 4C +28.07, we have studied the correlation between radio and $\gamma$-ray emission from this source during flares in its $\gamma$-ray light curve.  Below, we highlight the important aspects of our study.
 \begin{itemize}
 \item First broadband study ever done to understand the nature of the blazar 4C +28.07.
 \item Three major flares along with many sub-flares have been observed in decade long $\gamma$-ray light curve ($\sim$ 12 years).
 \item A significant correlation between $\gamma$-ray and radio light curves was observed with a time lag of $\sim$ 70 days and $\sim$ 135 days for Flare-A and Flare-C respectively. However, no significant correlation was seen for Flare-B.
 \item $\gamma$-ray production can happen efficiently by EC mechanism near the BLR region, while the time delay in radio emission indicates its emission region is separated by $\sim$ 9 pc and $\sim$ 4 pc from $\gamma$-ray emission region for Flare-A and Flare-C respectively.
 \item To account for the different emission zones of radio and $\gamma$-ray emission for Flare-A and Flare-C, two zone emission model is used for the broadband SED modeling under the leptonic scenario.
 \end{itemize}

 More correlation study with multi-wavelength emission can show how the multiple emission zones may evolve or disappear during the lifetime of AGNs. 
Blazar studies, in the past, have shown that in most of the sources single emission zone is sufficient enough to explain the broadband SED in the leptonic scenarios. There are also cases where the hadronic model (\citealt{B_ttcher_2009, B_ttcher_2013}) has been preferred to explain the production of simultaneous broadband emission. Having this kind of degeneracy in the model has always been the topic of investigation in order to draw any concrete conclusions. So, more and more broadband study of blazars can help us to understand the common features of AGNs, their emission mechanisms (leptonic and/or hadronic) and their emission zones (single/multiple). Our study suggests that though in FSRQs leptonic model is more preferred over the hadronic model, multiple emission zones may be required in some cases to explain the broadband SED.  

\section{Software and third party data repository citations} \label{sec:cite}
This work has made use of publicly available Fermi LAT data obtained from FSSC's website data server and provided by NASA Goddard Space Flight Center. This work has also made use of data, software/tools obtained from NASA High Energy Astrophysics Science Archive Research Center (HEASARC) developed by Smithsonian astrophysical Observatory (SAO) and the XRT Data Analysis Software (XRTDAS) developed by ASI Science Data Center, Italy. This research has made use of data from the OVRO 40-m monitoring program \citep{2011ApJS..194...29R} which is supported in part by NASA grants NNX08AW31G, NNX11A043G, and NNX14AQ89G and NSF grants AST-0808050 and AST-1109911. The archival data from Submil-limeter Array observatory has also been used in this study \citep{2007ASPC..375..234G}.  The Submillimeter Array is a joint project between the Smithsonian Astrophysical Observatoryand the Academia Sinica Institute of Astronomy and Astro-physics and is funded by the Smithsonian Institution and the Acedemia Sinica. The CSS survey \citep{2009ApJ...696..870D} is funded by the National Aeronautics and Space Administration under Grant No. NNG05GF22G issued through the Science Mission Directorate Near-Earth Objects Observations Program. The CRTS survey is supported by the U.S.~National Science Foundation under grants AST-0909182 and AST-1313422.

\textbf{Acknowledgements :}
 We thank the referee for helpful comments to improve this paper. A.K. Das thanks T. Ghosh, S.K. Mondal, Hemanth M., N.N. Patra for proof reading and fruitful discussions. R.P. acknowledges the support by the Polish Funding Agency National Science Centre, project 2017/26/A/ST9/00756 (MAESTRO 9), and MNiSW grant DIR/WK/2018/12. R.P also thanks to Swayamtrupta Panda for helping with Cartoon.

\software{ Fermitools (\url{https://fermi.gsfc.nasa.gov/ssc/data/analysis/scitools/}) \\ 
GAMERA \href{https://github.com/libgamera/GAMERA}{(https://github.com/libgamera/GAMERA)}
HEASARC \href{https://heasarc.gsfc.nasa.gov/docs/software/heasoft/download.html}{(https://heasarc.gsfc.nasa.gov/docs/software/heasoft/)}
XSPEC \href{https://heasarc.gsfc.nasa.gov/xanadu/xspec/}{(https://heasarc.gsfc.nasa.gov/xanadu/xspec/)}
XSELECT \href{https://heasarc.gsfc.nasa.gov/ftools/xselect/}{(https://heasarc.gsfc.nasa.gov/ftools/xselect/)}}



\bibliographystyle{plain}

\bibliography{sample63}{}
\bibliographystyle{aasjournal}

\begin{table}[t]
\caption{The value of peak time ($t_0$) \& peak flux ($F_0$) are given in column 2 \& column 3 respectively. column 4 \& column 5 represent the rising ($T_r$) \& decay time ($T_d$). Here, results are shown for 3 day binning light curve (Flare-A).} 
\label{tab:1}
\centering
\begin{tabular}{ccccc rrrr}   
\\
 & & Flare-A\\
\hline\hline                        
Peak & $t_0$ & $F_0$ & $T_r$ & $T_d$ \\ [0.8ex] 
& [MJD] & [10$^{-6}$ ph cm$^{-2}$ s$^{-1}$] & [day] & [day]\\
\hline
P1 &  55839.5 & 1.37$\pm$0.13 &  5.64$\pm$0.44 & 1.41$\pm$0.32 & \\
P2 &  55884.5 & 1.55$\pm$0.13 &  3.67$\pm$0.51 & 2.37$\pm$0.60 & \\
\hline                          
\end{tabular}
\end{table}

\begin{table}[t]
\caption{All the mentioned parameters are same as Table-\ref{tab:1}} 
\label{tab:2}
\centering
\begin{tabular}{ccccc rrrr}   
\\
& & Flare-B\\
\hline\hline
& & flare-B1\\
\hline\hline                         
Peak & $t_0$ & $F_0$ & $T_r$ & $T_d$ \\ [0.8ex]
& [MJD] & [10$^{-6}$ ph cm$^{-2}$ s$^{-1}$] & [day] & [day]\\
\hline
P1 &  57866.5 & 1.14$\pm$0.12 &  5.24$\pm$0.62 & 3.89$\pm$1.37 & \\
P2 &  57881.5 & 1.07$\pm$0.12 &  4.02$\pm$1.59 & 5.58$\pm$2.74 & \\
P3 &  57902.5 & 0.84$\pm$0.11 &  2.05$\pm$0.70 & 3.08$\pm$0.55 & \\
P4 &  57923.5 & 0.87$\pm$0.11 &  3.00$\pm$0.46 & 2.42$\pm$0.43 & \\
\hline                          
 & & flare-B2\\ 
\hline\hline 
P1 &  57983.5 & 1.04$\pm$0.10 &  5.22$\pm$1.16 & 1.48$\pm$0.47 & \\
P2 &  57989.5 & 1.21$\pm$0.11 &  2.77$\pm$0.40 & 1.17$\pm$0.28 & \\
\hline
\end{tabular}
\end{table}

\begin{table}[t]
\caption{All the mentioned parameters are same as Table-\ref{tab:1}} 
\label{tab:3}
\centering
\begin{tabular}{ccccc rrrr}   
\\
 & & Flare-C\\
\hline\hline                        
Peak & $t_0$ & $F_0$ & $T_r$ & $T_d$ \\ [0.8ex]
& [MJD] & [10$^{-6}$ ph cm$^{-2}$ s$^{-1}$] & [day] & [day]\\
\hline
P1 &  58372.5 & 1.95$\pm$0.57 &  5.15$\pm$0.45 & 11.23$\pm$0.81 & \\
P2 &  58402.5 & 1.79$\pm$0.18 &  4.10$\pm$0.61 & 3.93$\pm$0.76 & \\
\hline                          
\end{tabular}
\end{table}

\begin{table*}[h]
\caption{Result of $\gamma$-ray SEDs for Flare-A, which are fitted with different models: PL and LP (see text for more details). Column 1 represents the different periods of activity, column 2 and column 3 to column 5 represent the total Flux ($F_0$) during the activity and parameters of different models respectively. The goodness of fit ($log(\mathcal{L})$) is mentioned in column 6. Column 7 represents the difference in the goodness of fit compared to PL model.} 
\label{tab:4}
\centering
\begin{tabular}{ccccc rrrr}   
\hline\hline                        
& & & Powerlaw\\
\hline
Activity & $F_0$ & $\Gamma_{PL}$ & & & -log(Likelihood) \\ [1.5ex]
& [10$^{-6}$ ph cm$^{-2}$ s$^{-1}$] &  \\
\hline
Pre-flare &  0.28$\pm$0.01 & 2.29$\pm$0.04&  - & - & 195076.01 & - \\
Flare &  0.98$\pm$0.002 & 2.29$\pm$0.04 &  - & - & 201995.31 & - \\
Post-flare &  0.43$\pm$0.02 & 2.36$\pm$0.03 &  - & - & 228257.65 & - \\
\hline                          
 & & & Logparabola\\ 
\hline
Activity & $F_0$ & $\alpha$ & $\beta$ & - & -log(Likelihood) & $\Delta$log(Likelihood) \\ [1.5ex]
& [10$^{-6}$ ph cm$^{-2}$ s$^{-1}$] &  \\
\hline
Pre-flare &  0.27$\pm$0.02 & 2.20$\pm$0.07 &  0.05$\pm$0.02 & - & 195074.42 & -1.59 \\
Flare &  0.94$\pm$0.02 & 2.14$\pm$0.03 &  0.11$\pm$0.02 & - & 201969.99 & -25.32 \\
Post-flare &  0.42$\pm$0.02 & 2.28$\pm$0.05 &  0.05$\pm$0.02 & - & 228255.26 & -2.39 \\

\hline
\hline 
\end{tabular}
\end{table*}

\begin{table*}[h]
\caption{All the described parameters are same as Table-\ref{tab:4}, but for Flare-B.} 
\label{tab:5}
\centering
\begin{tabular}{ccccc rrrr}   
\hline\hline                        
& & & Powerlaw\\
\hline
Activity & $F_0$ & $\Gamma_{PL}$ & & & -log(Likelihood) \\ [1.5ex]
& [10$^{-6}$ ph cm$^{-2}$ s$^{-1}$] &  \\
\hline
Pre-flare &  0.18$\pm$0.01 & 2.18$\pm$0.04 &  - & - & 318041.29 & - \\
flare-B1 &  0.53$\pm$0.01 & 2.14$\pm$0.02 &  - & - & 277904.79 & - \\
flare-B2 &  0.55$\pm$0.02 & 2.10$\pm$0.02 &  - & - & 163365.33 & - \\
Post-flare &  0.20$\pm$0.01 & 2.43$\pm$0.06 &  - & - & 196929.05 & - \\
\hline                          
 & & & Logparabola\\ 
\hline
Activity & $F_0$ & $\alpha$ & $\beta$ & - & -log(Likelihood) & $\Delta$log(Likelihood) \\ [1.5ex]
& [10$^{-6}$ ph cm$^{-2}$ s$^{-1}$] &  \\
\hline
Pre-flare &  0.17$\pm$0.01 & 2.02$\pm$0.07 &  0.07$\pm$0.03 & - & 318036.96 & -3.33 \\
flare-B1 &  0.50$\pm$0.01 & 1.97$\pm$0.04 &  0.08$\pm$0.01 & - & 277887.31 & -17.48 \\
flare-B2 &  0.51$\pm$0.02 & 1.87$\pm$0.04 &  0.12$\pm$0.02 & - & 163345.13 & -20.02 \\
Post-flare &  0.19$\pm$0.01 & 2.32$\pm$0.09 &  0.07$\pm$0.04 & - & 196927.64 & -1.41 \\

\hline
\hline 
\end{tabular}
\end{table*}

\begin{table*}[h]
\caption{All the described parameters are same as Table-\ref{tab:4}, but fot Flare-C.} 
\label{tab:6}
\centering
\begin{tabular}{ccccc rrrr}   
\hline\hline                        
& & & Powerlaw\\
\hline
Activity & $F_0$ & $\Gamma_{PL}$ & & & -log(Likelihood) \\ [1.5ex]
& [10$^{-6}$ ph cm$^{-2}$ s$^{-1}$] &  \\
\hline
Flare &  0.68$\pm$0.02 & 2.10$\pm$0.02 &  - & - & 271738.69 & - \\
Post-flare &  0.31$\pm$0.01 & 2.16$\pm$0.02 &  - & - & 334160.84 & - \\
\hline                          
 & & & Logparabola\\ 
\hline
Activity & $F_0$ & $\alpha$ & $\beta$ & - & -log(Likelihood) & $\Delta$log(Likelihood) \\ [1.5ex]
& [10$^{-6}$ ph cm$^{-2}$ s$^{-1}$] &  \\
\hline
Flare &  0.64$\pm$0.02 & 1.95$\pm$0.03 &  0.07$\pm$0.01 & - & 271719.74 & -18.95 \\
Post-flare &  0.30$\pm$0.01 & 2.03$\pm$0.05 &  0.06$\pm$0.02 & - & 334154.53 & -6.31 \\

\hline
\hline 
\end{tabular}
\end{table*}

\begin{table*}[h]
\caption{Results of reduced-$\chi^2$ value (column 2) for different spectral models (Powerlaw and Logparabola). Column 1 represents different flares activity.} 
\label{tab:7}
\centering
\begin{tabular}{c|cccc rrrr}   
\hline\hline                        
Activity & & Reduced-$\chi^2$ \\ [0.8ex]
\hline
 & Powerlaw & Log-parabola \\
\hline
Flare-A \\
\hline
flare & 16.75 & 2.02 \\
\hline
Flare-B \\
\hline
flare-B1 & 8.03 & 1.71 \\
flare-B2 & 8.48 & 1.84 \\
\hline
Flare-C \\
\hline
flare & 9.28 & 2.74 \\
\hline\hline
\end{tabular}
\end{table*}

\begin{table*}
\caption{Results of multi-wavelength SED modelling. 1st column represents the study of different periods. Time span of different activities are given in last column (see text for more details).} 
\label{tab:8}
\centering
\begin{tabular}{ccccc rrrr}   
\hline\hline
Activity & Parameters & Symbol & BLOB-I & BLOB-II & Time duration  \\
\hline\hline
& Temperature of BLR region & $T^\prime_{BLR}$ & 1.0$\times10^{4}$ K & $\times$  \\
& Phonton density of BLR region & $U^\prime_{BLR}$ & 4.28 erg/$cm^3$  & $\times$  \\
& Temperature of Disk & $T^\prime_{Disk}$ & 6.0$\times10^{4}$ K & $\times$ \\
& Phonton density of Disk & $U^\prime_{Disk}$ & 7.0$\times10^{-8}$ erg/$cm^3$ & $\times$ \\
& Size of the emission region & R & 2.0$\times10^{16}$ cm & 6.5$\times10^{17}$ cm \\
& Doppler factor of emission region & $\delta$ & 25 & 25  \\
& Lorentz factor of the emission region & $\Gamma$ & 12.7 & 12.7 \\ 
\hline\hline
& Spectral index of injected electron spectrum  & $\alpha$ & 1.90 & 1.82  \\
& Curvature index of electron spectrum & $\beta$ & 0.15 & 0.15 \\
& Min energy of injected electrons & $e_{min}$ & 3.58 MeV & 148 MeV  \\
flare-A & Max enrgy of injected electrons & $e_{max}$ & 4854 MeV & 3321 MeV & 71 days  \\
& Magnetic field & B & 1.20 G & 0.25 G \\
& Power in the injected electrons & $P_e$ & 0.71$\times10^{46}$ erg/sec & 5.86$\times10^{46}$ erg/sec  \\
& Power in the magnetic field & $P_B$ & 0.03$\times10^{46}$ erg/sec & 1.59$\times10^{46}$ erg/sec  \\
& Total required jet power & $P_{tot}$ & 1.61$\times10^{46}$ erg/sec & 8.03$\times10^{46}$ erg/sec\\
\hline
& Spectral index of injected electron spectrum  & $\alpha$ & 1.77 & 2.01 \\
& Curvature index of electron spectrum & $\beta$ & 0.11 & 0.06 \\
& Min energy of injected electrons & $e_{min}$ & 2.4 MeV & 135.9 MeV \\
flare-B2 & Max enrgy of injected electrons & $e_{max}$ & 7562.8 MeV & 3321.5 MeV & 42 days \\
& Magnetic field & B & 2.1 G & 0.14 G \\
& Power in the injected electrons & $P_e$ & 0.37$\times10^{46}$ erg/sec & 6.59$\times10^{46}$ erg/sec \\
& Power in the magnetic field & $P_B$ & 0.11$\times10^{46}$ erg/sec & 0.50$\times10^{46}$ erg/sec \\
& Total required jet power & $P_{tot}$ & 0.88$\times10^{46}$ erg/sec & 7.79$\times10^{46}$ erg/sec \\
\hline
& Spectral index of injected electron spectrum & $\alpha$ & 1.80 & 2.01  \\
& Curvature index of electron spectrum & $\beta$ & 0.10 & 0.06 \\
& Min energy of injected electrons & $e_{min}$ & 2.45 Mev & 148.2 Mev  \\
flare-C & Max enrgy of injected electrons & $e_{max}$ & 8176.0 MeV & 3321.5 MeV & 66 days \\
& Magnetic field & B & 2.3 G & 0.15 G \\
& Power in the injected electrons & $P_e$ & 0.44$\times10^{46}$ erg/sec & 6.03$\times10^{46}$ erg/sec \\
& Power in the magnetic field & $P_B$ & 0.13$\times10^{46}$ erg/sec & 0.57$\times10^{46}$ erg/sec  \\
& Total required jet power & $P_{tot}$ & 1.06$\times10^{46}$ erg/sec & 7.20$\times10^{46}$ erg/sec\\
\hline
& Spectral index of injected electron spectrum & $\alpha$ & 1.90 & 2.01 \\
& Curvature index of electron spectrum & $\beta$ & 0.08 & 0.06 \\
& Min energy of injected electrons & $e_{min}$ & 1.84 MeV & 163.5 MeV  \\
Quiescent State & Max enrgy of injected electrons & $e_{max}$ & 5110.0 MeV & 3066.0 MeV & 3 days \\
& Magnetic field & B & 1.7 G & 0.10 G\\
& Power in the injected electrons & $P_e$ & 1.46$\times10^{46}$ erg/sec & 4.77$\times10^{46}$ erg/sec  \\
& Power in the magnetic field & $P_B$ & 0.07$\times10^{46}$ erg/sec & 0.25$\times10^{46}$ erg/sec  \\
& Total required jet power & $P_{tot}$ & 1.81$\times10^{46}$ erg/sec & 5.48$\times10^{46}$ erg/sec\\

\hline\hline
\end{tabular}
\end{table*}

\end{document}